\documentclass{ar1colS2O}

\usepackage[comma]{natbib}
\usepackage{url}
\usepackage{amsmath,amssymb}
\usepackage{subcaption}
\usepackage{xcolor}
\usepackage[most]{tcolorbox}
\definecolor{arblue}{RGB}{20, 50, 110} 

\DeclareRobustCommand{\ion}[2]{%
\relax\ifmmode
\ifx\testbx\f@series
{\mathbf{#1\thinspace\mathsc{#2}}}\else
{\mathrm{#1\thinspace\mathsc{#2}}}\fi
\else\textup{#1\thinspace\textsc{#2}}%
\fi}

\def\Ledd{L_{\rm{Edd}}}
\def\Medd{\dot{M}_{\rm{Edd}}}
\def\msun{M_{\odot}}
\def\mbh{M_{\rm BH}}
\newcommand{\SFinf}{\mathrm{SF}_\infty}

\jname{Xxxx. Xxx. Xxx. Xxx.}
\jvol{AA}
\jyear{YYYY}
\doi{10.1146/((please add article doi))}

\begin{document}

\markboth{Jiang \& Greene}{Super-Eddington Accretion}

\title{Super-Eddington Accretion onto Compact Objects: Theory Meets Observation}

\author{Yan-Fei Jiang,$^1$ and Jenny E. Greene,$^2$
\affil{$^1$Center for Computational Astrophysics, Flatiron Institute, New York, NY 10010, USA}
\affil{$^2$Department of Astrophysical Sciences, Princeton University, 4 Ivy Lane, Princeton, NJ 08544, USA}}

\begin{abstract}
\setlength{\linewidth}{\hsize}
\makeatletter
\@rightskip=18pt
\rightskip=\@rightskip
\makeatother
We review candidate super-Eddington populations together with semi-analytic models and radiation magnetohydrodynamic simulations, with the following conclusions: 
$\bullet$ Likely super-Eddington sources over a range of black hole mass show similar properties, including outflow on a range of scales, hard X-ray weakness, decreased emission from jets, and suppressed optical/UV variability. 
$\bullet$ Theoretical treatment of super-Eddington accretion predicts similar behavior across a range of methods and initial conditions, although interesting differences remain, particularly in the importance of magnetic fields. Numerical simulations produce a puffed-up inner disk with a funnel that grows narrower at higher mass accretion rate. Radiative efficiency declines with accretion rate in thermal-pressure--supported disks. 
$\bullet$ Ubiquitous outflow is likely an important ingredient in galaxy feedback at high Eddington ratios, as wind mechanical power increases with accretion rate and can rival jet power. Wind kinetic power depends far less on uncertain black hole spin and magnetic fields. $\bullet$ Observational classes of objects can be sorted in a two-dimensional parameter space of Eddington ratio and orientation. 
Observations down the funnel observe rapid X-ray variability, while observations through the wind will see suppressed X-ray emission. At the highest Eddington ratios, we envision a regime where the photosphere extends to $r > 10^6 \, r_g$, and the observer sees only thermalized emission covering most of the sky, speculatively corresponding to super-soft sources and Little Red Dots.
\end{abstract}

\maketitle

\tableofcontents

\section{INTRODUCTION}
The luminosity at which the radiation force for gas balances gravity is the Eddington luminosity $\Ledd\equiv 4\pi G Mc/\kappa=1.25\times 10^{38}\left(M/\msun\right)/(\kappa/0.4\ {\rm cm}^2 {\rm g}^{-1})$ erg/s. If the Eddington luminosity is produced by gas accreted onto a black hole, the associated Eddington mass accretion rate is $\Medd\equiv \Ledd/(0.1 c^2)=2.21 \left(M/10^8\msun\right)\ \msun/{\rm yr}$. As an analogy to the Eddington limit of stellar structure \citep{Eddington1926}, it has long been believed that the accretion rate onto black holes also cannot exceed the Eddington limit. Statistically, large quasar samples with estimated black hole masses and bolometric luminosity also show a sharp scarcity of objects that exceed the Eddington luminosity \citep[e.g.,][]{Shen+2011}. However, this does not prove that super-Eddington accretion is not possible. It may also reflect that a super-Eddington accretion state can only be sustained for short periods of time due to interactions between the central engine and the surrounding environment. Therefore, super-Eddington sources are less frequently observed compared with their long lived sub-Eddington counterparts, but could still have an important role to play in the growth of black holes and galaxies.



We have long known that super-Eddington sources exist in nature. Perhaps the strongest example is SS 433 \citep{Fabrika2004}, which produces a jet and optically thick wind. This source belongs to a special class of X-ray binaries: ultraluminous X-ray sources (ULXs) \citep{FengSoria2011}. 
The super-Eddington picture is also becoming more plausiable as 
coherent pulsations from M82 X-2 were discovered, showing that accretion onto a neutron star can produce apparent luminosity far above the canonical Eddington limit \citep{Bachetti+2014}. These objects have special spectral properties compared with typical X-ray binaries. There are also a small fraction of active galactic nuclei (AGNs) showing unique spectral properties associated with super-Eddington accretion, such as narrow H$\beta$ lines, strong \ion{Fe}{II} emission, weak [\ion{O}{III}] lines and weak X-ray emission, as well as soft X-ray spectra. However, without clear predictions of the signatures of super-Eddington accretion, it is hard to claim that these outliers are radiating above the Eddington limit.

Theoretically, since the standard thin disk model was proposed, it has been understood that the accretion disk behaves differently when the accretion rate exceeds the Eddington limit, as a radiation-driven wind may emerge \citep{ShakuraSunyaev1973}. The development of theoretical models for super-Eddington accretion primarily focus on checking the following aspects of the Eddington argument: spherical symmetry, hydrostatic balance, homogeneity, and energy transport, as well as the effects of magnetic fields and opacity. Extensions of the standard thin disk model to the super-Eddington regime have been worked out by including the effects of advective energy transport and radial pressure support against gravity \citep{Begelman1978,Abramowicz+1988,AbramowiczFragile2013}. Anisotropy is also proposed as one of the key properties for super-Eddington accretion. Mass can accrete near the disk midplane with small radial radiative luminosity while most of the photons escape in a direction perpendicular to the disk \citep{Jaroszynski+1980,Abramowicz+1980,King2009}. Large density fluctuations at the scale of the photon mean free path can allow photons to escape through low density channels without pushing away most of the mass. This is in close analogy to the clumpy wind in massive stars \citep{Shaviv1998} and has also been suggested as a way to generate super-Eddington luminosity in accreting systems \citep{Begelman2002}. 

Advancements in computational astrophysics, both in computing power and development of numerical algorithms for radiation magneto-hydrodynamics, make it possible to explore the multi-dimensional nature of super-Eddington accretion disks. Early two-dimensional radiation hydrodynamic simulations using the flux limited diffusion approximation for radiation transport and the $\alpha$ prescription for angular momentum transport show that super-Eddington accretion is indeed possible with a structure roughly consistent with theoretical expectations \citep{Eggum+1985,Ohsuga+2005,OhsugaMineshige2007}. Self-consistent magneto-hydrodynamic turbulence in three dimensions, more accurate solvers for radiation transport and realistic atomic physics for radiation and gas coupling, along with general relativity, have been added to improve the accuracy of the simulations 
\citep{Sadowski+2014,Jiang+2014a,McKinney+2014,TakahashiOhsuga2015,Jiang+2019,Zhang+2025,Fragile+2025}. These simulations show a more complicated picture than the original theoretical models. The properties of the resulting super-Eddington disks may depend on how the disk is fed at large scales. Connecting the simulation results to observational properties of super-Eddington sources becomes an important way to distinguish different models and constrain the free parameters used in these simulations.

Beyond accretion physics, there are many implications if super-Eddington accretion is common. First of all, the prevalence of super-Eddington growth has profound implications for the birth and early growth of massive black holes. We observe $M_{\rm BH} \approx 10^9~\msun\ $growing black holes only $\sim 660-670$ Myr after the Big Bang \citep[e.g.,][ and references therein]{Fan+2023}. If growth is capped at the Eddington limit, then their progenitors must start `heavy', at least $M_{\rm BH} \approx 10^4~\msun$, in order to reach a billion suns in the limited available time \citep[e.g.,][]{Turner:1991,HaimanLoeb:2001}. In contrast, with super-Eddington growth as a possibility, the initial seed mass can take a much broader range of values \citep[e.g.,][]{Inayoshi+2020,Trinca+2024}. Even starting from $10^3-10^5 \, M_{\odot}$ seed black holes, the black hole needs to grow semi-continuously at the Eddington rate to achieve a billion suns in a few hundred million years \citep[e.g.,][]{Johnson+2012,Volonteri+2015}, while short episodes of super-Eddington growth would alleviate this problem.

If black holes can grow above the Eddington limit, then it is a high priority to find them, count them, and understand the prevalence of this growth mode as a function of cosmic time. The emergent spectra of accreting black holes are certainly a function of accretion rate, although observationally it can be very tricky to identify super-Eddington sources given fundamental uncertainties in the determination of both black hole mass and bolometric luminosity (see \S \ref{sec:obs_summary}). 
There are special sub-classes of sources that are likely to have super-Eddington accretion rates (see \S \ref{sec:sub_classes}). All sources are very luminous given their likely black hole mass, but some are luminous in the UV and show specific line properties in the UV/optical region, while others are heavily reddened and MIR luminous. Some show excess variability in the X-ray, while others are profoundly X-ray quiet. 

The only way to unravel this confusing picture of super-Eddington sources is to connect theoretical progress over the past decades with observational phenomenology, which is the primary goal of this review. There have been numerous self-consistent radiation magneto-hydrodynamic simulations that show what the super-Eddington accretion disks should look like (see \S \ref{sec:simulation}). 
We will show that the core properties associated with Eddington and super-Eddington accretion can be understood in the context of changes in accretion disk structures with accretion rates as well as the dependence on viewing angles (see \S \ref{sec:theo_summary}). 

One common feature across nearly all sources that are identified as super-Eddington is outflow. There are many important questions associated with these outflows. If they are launched in the inner accretion flow, then how are they driven outward and what fraction of the bolometric luminosity do they carry? Do these winds couple efficiently to gas in the larger galaxy to shorten the accretion lifetime? We will investigate theoretical expectations of the link between accretion rate and outflow power and then evaluate the extent to which observations can test these predictions (\S \ref{sec:theo_summary}). 

We note there have been multiple recent reviews on topics related to super-Eddington accretion, such as theoretical models for super-Eddington accretion disks \citep{AbramowiczFragile2013}, outflow and feedback 
\citep{Fabian2012, KingPounds2015, Veilleux:2020}, growth of black holes in the early universe \citep{Inayoshi+2020,Fan+2023}, properties of ULX \citep{FengSoria2011,KingLasotaMiddleton2023,Kaaret2017}, numerical simulations of super-Eddington accretion flows \citep{JiangDai2025}, super-Eddington accretion in high-z universe \citep{Mayer2019}, and observational properties of super-Eddington quasars \citep{Marziani+2025}. The primary goal of this review is to join  theoretical and the observational perspectives to propose a picture of super-Eddington accretion.

\section{Observational Properties of Systems with Accretion Rate Near or Above the Eddington limit}
\label{sec:obs_summary}

Before we can determine how black hole properties (spectral energy distributions, outflows, variability) depend on accretion rate, we must have a working understanding of the black hole mass and the bolometric luminosity. Thus we provide a very brief summary of methods to determine black hole mass and $L_{\rm bol}$, and systematic uncertainties in these determinations.

\begin{marginnote}[]
	\entry{Reverberation mapping}{The stochastic variability of AGN lightcurves in different wavelength bands is known to correlate with variability in the broad lines, showing a lag of time-scale $\tau$. 
	This lag is thought to arise as the variable continuum photoionizes the gas with a delay corresponding to the light travel time from the central accretion disk to the broad line region. The lag $\tau$ gives the emissivity-weighted radius of the broad line region. Combined with the width of the lines, the black hole mass can be estimated assuming the velocity that broadens the lines is virialized with the potential of the black hole. }
\end{marginnote}



\subsection{Measuring black hole mass and eddington ratio}


%

The most reliable black hole masses for actively accreting supermassive black holes come from either reverberation mapping or for a small number of very luminous targets, from direct dynamical measurements using the broad-line region \citep[BLR; e.g.,][ discussed below]{Gravity+2024a}. For the vast majority of quasars, reverberation mapping is not available. However, there is a correlation between the observed AGN luminosity and the radius of the broad-line region, which is calibrated at low redshift \citep[e.g.,][]{Kaspi:2000,Bentz+2013}. These radius-luminosity relations are then applied to derive an average radius for sources as a function of their luminosity. The virial assumption with an average geometric $f$ value yields a black hole mass, typically called a `single-epoch' black hole mass \citep{Vestergaard+2002}. 

The single-epoch black hole masses are really order-of-magnitude estimates with significant ($\sim 0.5$~dex) scatter and likely systematic shifts with black hole mass, Eddington ratio, and potentially redshift that are not easily quantified. In particular, the reverberation mapped samples are heavily weighted to low redshift, although they extend to $z = 4.5$ \citep[e.g.,][]{Malik+2023,Shen:2024}. Thus, there are few lag measurements for the most luminous quasars at Cosmic Noon. Some very reassuring tests have come from near infrared interferometry with the upgraded GRAVITY+ instrument \citep{Abuter+2022}. GRAVITY+ can measure spatially resolved kinematics across the broad-line regions of a handful of AGNs from the local universe to redshift $\sim 4$ \citep[e.g.][]{Gravity+2024a,Gravity+2026}. By combining light from the four 8 meter Unit Telescopes, the GRAVITY+ instrument is able to measure the wavelength-dependent photo-center shift of a broad emission line across its velocity profile. Combined with specific modeling of the broad line region, these measurements can constrain the ordered rotation of the broad line gas and therefore yield a dynamical mass of the central black hole. The Keplerian rotation model for the line emitting gas works very well for the low luminosity local AGNs and GRAVITY+ gets consistent size-luminosity relationship as obtained by reverberation mapping \citep{Gravity+2024a}.

However, the same model cannot be used to describe the properties of the observed H$\beta$ and H$\gamma$ lines by GRAVITY+ for the quasar SMSS J052915.80-435152.0 at redshift $\approx 3.98$. A prominent blue wing in the line profile requires that more than $80\%$ of the line emission is produced by an outflow-dominated component \citep{Gravity+2026}, while the remaining emission can be produced by a rotating disk. The black hole mass obtained by this model is much smaller than the estimated value based on the single epoch approach. Therefore, this is an object with a super-Eddington accretion rate, which may explain the significantly different dynamical structures in the broad line region. The outflow rate estimated based on the forward modeling is only $\approx 1\%$ of the mass accretion rate, which can be subject to significant uncertainties in the forward modeling to separate the outflow and rotating components based on the observed line profiles. 


We also see evidence that the black hole mass estimators depend on accretion rate. As explained in more detail below (\S \ref{sec:variability}), at fixed luminosity, the measured lag between the broad line region and the continuum variability actually decreases as the Eddington ratio increases. Furthermore, there is tentative evidence from spatially resolved reverberation mapping that the geometric $f$-factor changes with $L_{\rm bol}/L_{\rm Edd}$, which leads to systematic bias in the black hole mass inference \citep[e.g.,][]{Villafana+2022,Linzer+2022}. We must wait until we have more direct calibration of black hole masses at high luminosity from higher-redshift reverberation mapping campaigns \citep[e.g.,][]{Shen:2024} along with more dynamical measurements from GRAVITY+ and from stars or gas using \emph{JWST} \citep{Newman+2026} or E-ELT \citep{Nguyen+2026}.

The super-Eddington Accreting Massive Black Holes 
(SEAMBH) project presents a dedicated optical RM campaign targeting AGNs accreting at or above the Eddington rate (mostly NLS1s) \citep{Du+2014}. The measured H$\beta$ lags for these sources are systematically below the predictions of the scaling relations as found by \cite{Bentz+2013} with the measured $5100$~\AA\ luminosity for these sources \citep{Du+2015}. The differences between the measured and predicted lags are also found to be tightly correlated with the Eddington ratio. Sources with larger Eddington ratios show bigger deficits. A revised tight correlation can be established for these sources if a separate term related to the Eddington ratio (see \S \ref{sec:EV1}) is added \citep{Du+2016,DuWang2019}. It has been suggested that the ionizing photons emitted from the central region may be shadowed by the thick disk itself when the accretion rate is above the Eddington limit, which causes the broad lines to emerge at a radius closer to the black hole \citep{Wang+2014,Wu+2025}. However, detailed photo-ionization studies show that simply reducing the ionization flux is not enough to explain the deficit in the H$\beta$ lags. It is possible that increased gas density within the broad line region also contributes to the difference, which may be more prevalent for smaller mass black holes or higher mass accretion rate \citep{Wu+2026}.

\begin{figure}[htbp]
	\centering
	\includegraphics[width=\linewidth]{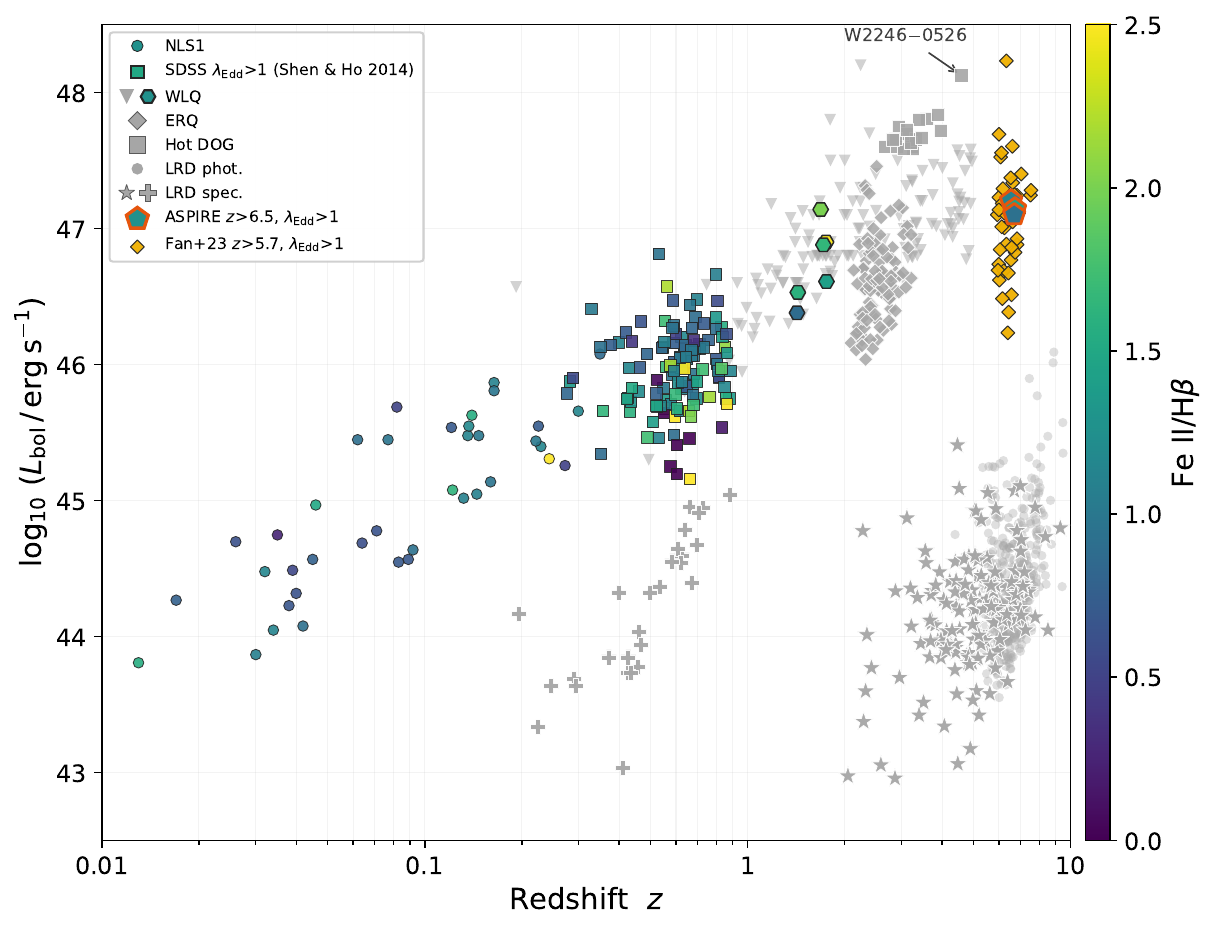}
	\caption{Bolometric luminosity versus redshift for the high-accretion and
	extreme AGN populations, after the class-specific bolometric corrections
	summarized in Table~\ref{tab:zLsamples}. Marker \emph{shape} encodes the
	class or sample, while point \emph{color} encodes the Eigenvector-1
   accretion-rate diagnostic $R_{\rm FeII}$
    wherever a direct
	per-object measurement exists. Objects with no such measurement are drawn
	in flat grey.
	\emph{Base populations:} NLS1s \citep{Grupe+2010}; weak-line quasars
	(WLQs; \citealt{Leighly+2007,Diamond-Stanic+2009,Wu+2011,Wu+2012,MeusingerBalanfkan2014,Luo+2015, Plotkin+2015});
	extreme red quasars (ERQs; \citealt{Hamann+2017}); hot dust obscured galaxies (Hot DOGs; \citealt{Tsai+2015}); and the
	photometric LRDs \citep{Akins+2025}, the 146 spectroscopic LRDs of
	\citet{deGraaff+2026} at $2.0\!<\!z\!<\!9.3$, and the 27 low-redshift
	DESI DR1 LRDs of \citet{Lin+2026} at $z\!=\!0.2$--0.9.
	Every object plotted here, with its redshift, observed luminosity and
	bolometric luminosity, is available as a machine-readable table in the
	Supplemental Material.
	\emph{Super-Eddington samples highlighted here:} the $z>6.5$ JWST/ASPIRE
	quasars \citep{Yang+2023}, of which only the three with
	$\lambda_{\rm Edd}>1$ are plotted (orange-edged pentagons); the 40 $z>5.7$
	quasars from the \citet{Fan+2023} compilation with a Mg\,{\sc ii}-based
	$\lambda_{\rm Edd}>1$ (amber diamonds); and the 120 low-redshift SDSS
	quasars with a broad-H$\beta$ EV1 measurement and $\lambda_{\rm Edd}>1$
	from \citet{ShenHo2014} (squares).
  }
	\label{fig:redshift_lum}
\end{figure}

The final step required to estimate $L_{\rm bol}/L_{\rm Edd}$ is an estimate of the bolometric luminosity. Typically, full spectral energy distributions are not available, so an average bolometric correction is often assumed \citep[e.g.,][]{Elvis+1994,Richards+2006}. There is significant scatter in these relations that is hard to quantify. However, more concerning is the clear evidence that the spectral energy distributions of AGN are functions of Eddington ratio \citep[e.g.,][]{Ho:2008,VasudevanFabian2009}. Typically we do not have complete information for a given source, and without an isotropic and dust-insenstitive tracer like mid-infrared or X-ray luminosity, it can be very difficult to even estimate the bolometric luminosity. In \S \ref{sec:spec_model}, we will discuss our current theoretical understanding of how the SED will change with accretion rate.

In summary, it is clear that standard scaling relations to determine bolometric luminosity and black hole mass do depend on the Eddington ratio, causing systematic bias in estimates that ignore the accretion rate. Nevertheless, we will show below that it is possible to identify credible examples of AGN accreting at super-Eddington rates, even with these systemic uncertainties.

\subsection{Properties that depend on accretion rate}

There is substantial evidence that basic properties of AGN depend on the accretion rate. To establish baseline trends, we start with a short discussion of various properties (SED and variability) that are known to depend on luminosity and/or Eddington ratio. A summary of the different types of objects that we will discuss is presented in Figure \ref{fig:redshift_lum}, and Table \ref{tab:zLsamples}.

\subsubsection{Eigenvector 1}
\label{sec:EV1}

AGNs that are uniformly selected based on a blue color in the UV/optical region tend to be quite homogeneous in terms of their spectral shape, broader spectral energy distribution, and line properties \citep[e.g.,][]{SternLaor:2012,Fan+2023}. However, it is possible to find families of optical spectroscopic properties that seem to vary as a function of black hole mass and luminosity. An early comprehensive study of this kind was performed by \citet{BorosonGreen1992}, who performed a parameter reduction on a large sample of Palomar Green AGN \citep{SchmidtGreen1983}. A strikingly small number of eigenspectra could describe the full space, and they found that the so-called ``Eigenvector 1'' consisted of trends between narrow [\ion{O}{iii}]/H$\beta$, the strength of the broad \ion{Fe}{ii} emission, and the width of the broad H$\beta$ line. Even in that initial paper, the authors speculated that Eddington ratio is likely the driving factor behind the Eigenvector 1 trends. The relatively narrow H$\beta$ line can be understood as arising from lower-mass black holes at fixed luminosity. As far as we know, however, there is not a complete understanding of the trend towards lower [\ion{O}{iii}] and stronger FeII at higher Eddington ratio.

Subsequent studies of eigenvector 1 extended this analysis in a number of ways, both by refining the properties explored in the optical \citep[e.g.,][]{Marziani+2001} and by incorporating multi-wavelength information. We refer the interested reader to a comprehensive recent review by \citet{ShenHo2014}. They show that red mid-infrared color and weaker X-ray emission with a softer slope are both associated with the high-Eddington end of eigenvector 1. \citet{ShenHo2014} find the following trends with increasing eigenvector 1: (a) increasing soft X-ray photon index; (b) redder mid-infrared to optical color; (c) weaker high-ionization lines; (d) increasing radio-quiet fraction \citep{Sikora+2007}; and (e) decreasing optical variability. 

\subsubsection{Variability}
\label{sec:variability}

As mentioned in \S \ref{sec:EV1} above, the variability properties of accreting black holes are a function of Eddington ratio. Here we discuss more detailed trends in both the X-ray and UV/optical bands. 

{\it X-ray Variability.} X-ray data collected by ROSAT and \emph{Advanced Satellite for Cosmology and Astrophysics} (ASCA) \citep[e.g.,][]{Boller+1996} shows that high EV1 sources can show much larger and more rapid X-ray variability than typical AGN, with an anti-correlation between excess X-ray variability and the FWHM of the H$\beta$ line. As an extreme example, the flux in the $0.2-2$ keV band of ROSAT data for the high EV1 galaxy \emph{IRAS 13224-3809} can change by a factor of 57 in just two days \citep{Boller+1997}, while significant variations in the sub-day timescales are also common in many high EV1 galaxies \citep{Brandt+1999,Miniutti+2009}. Statistically, the high frequency power spectrum densities (PSDs) of Seyferts are usually described by power laws with $\mathcal{P}(f)\propto f^{-\gamma}, \gamma\approx 1-2$, where $\mathcal{P}(f)$ is the variability power at frequency $f$ \citep[e.g.,][]{Uttley+2002}. The power-law slope typically changes from $\gamma=1$ at long time scales to a steeper slope $\gamma=2$ at shorter timescales. The timescales, $T_B$, where the power-law variability power spectrum breaks for X-rays are found to correlate with black hole mass and mass accretion rate for both stellar mass and supermassive black holes as \citep{McHardy+2006} 
\begin{eqnarray}
    \log \left(T_B/{\rm days}\right)=2.1\log\left(\mbh/10^6\msun\right)-0.98\log\left( L_{\rm{Bol}}/10^{44} \rm{erg/s}\right)
    -2.32.
\end{eqnarray}
This is consistent with many other studies of X-ray excess variance of high EV1 sources \citep{Vaughan+2003,Ponti+2012}. On variability timescales of days to months, which are typically above the PSD breaks, the variability amplitude of the soft X-ray photons ($\lesssim 2$ keV) is typically suppressed while the hard X-rays can still show strong variability as found for Mrk 766 \citep{Vaughan+2003b}, 1H 0707-495 and IRAS 13224-3809 \citep{Kara+2013}. In addition to changes in amplitude between the hard and soft band, there are frequency-dependent changes as well, such that in some sources the soft X-rays lag the hard X-rays at high frequency, but lead the hard X-rays at low frequency \citep[e.g.,][]{Jin+2017b}. Apparently the hard and soft-X-ray photons are produced in different spatial locations. The observation that they show correlation likely arises from a combination of reverberation and accretion. In section \ref{sec:sim_spec}, we will describe a simulation result that explains the origin of soft X-ray photons compared with the hard X-rays produced by the corona.

\begin{figure}[htbp]
	\centering
		\includegraphics[width=\linewidth]{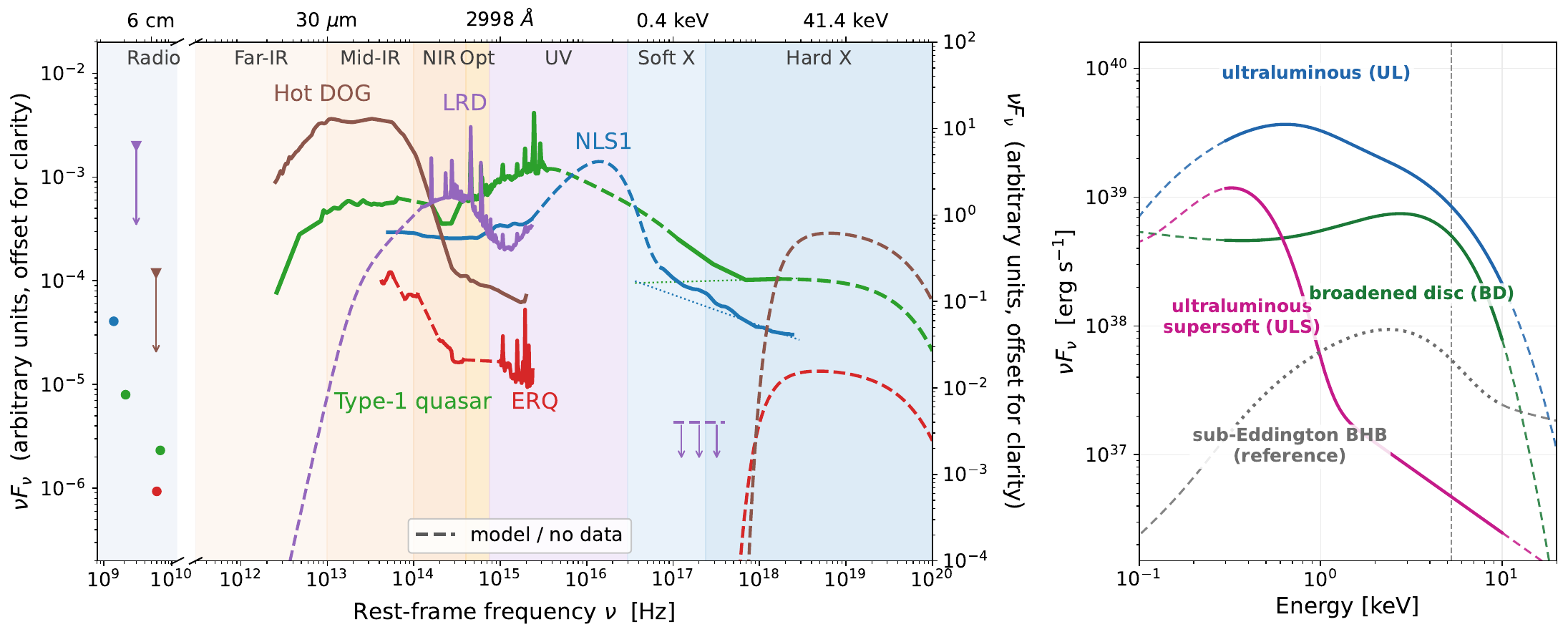}
		\caption{Spectral energy distributions ($\nu F_\nu$) of super-Eddington AGN
		(\emph{left}) and ultraluminous X-ray sources (\emph{right}). In both panels,
		\emph{solid marks measured data; dashed marks a model, or a range in which
		nothing has been measured.}
		\emph{Left:} broadband SEDs, radio to hard X-ray, of a standard type-1 quasar
		compared with the extreme AGN populations discussed here. Curves are
		normalized at rest-frame 1\,$\mu$m and offset for clarity. The type-1
		reference is the median SED of the 27 radio-quiet quasars of
		\citet{Shang+2011}, with the composites of \citet{VandenBerk+2001} and
		\citet{Glikman+2006} shortward of 0.7\,$\mu$m. The unobservable EUV
		(850\,\AA--0.5\,keV) is bridged following \citet{KubotaDone2018}. Dotted
		lines are the hard-X-ray power laws extrapolated down, so the soft excess is
		the flux above them (a factor 2.0 at 0.49\,keV for the type-1). NLS1s are
		the median of the 42 objects with FWHM(H$\beta$)$<$2000\,km\,s$^{-1}$ in
		the \emph{Swift} survey of \citet{Grupe+2010}, extended into the infrared
		using \citet{Shen+2011}. Hot DOGs are the median of 20 WISE ELIRGs
		\citep{Tsai+2015}. The ERQ curve is based on the BOSS spectra of all 97
		core ERQs of \citet{Hamann+2017} in the rest-UV (1260--2990\,\AA), and its
		near- to mid-infrared (0.80--7.1\,$\mu$m) is stacked from those same
		objects' AllWISE $W1$--$W4$. The solid LRD V-shape (rest
		$0.13$--$2.4\,\mu$m) is the median-stacked SED of the photometric
		COSMOS-Web LRDs of \citet{Akins+2025} and the spectroscopic sample used in
		\citet{deGraaff+2026}. The dashed mid- and far-infrared follows the MIRI
		detections and ALMA/\emph{Herschel} non-detections of \citet{Setton+2025}
		and \citet{Casey+2025}, and the purple arrows near 1\,keV mark the LRD
		soft-X-ray upper limit. Radio measurements are too sparse for curves and
		are shown as points and upper limits
		\citep{Rakshit+2017,Hwang+2018,Frey+2016,Perger+2025}.
		\emph{Right:} de-absorbed X-ray spectra of ULXs, each class at its own
		luminosity. Solid lines span the $0.3-10$ keV band in which the spectra
		were fitted, and dashed lines extrapolate the best-fitting models. The
		ultraluminous (UL, blue) and broadened disk (BD, green) curves are the
		medians of the multi-color disk plus power-law fits to 88
		\emph{XMM-Newton} observations of 20 ULXs by \citet{Sutton2013}, with the
		hard and soft ultraluminous sub-classes merged into UL. The ultraluminous
		supersoft (ULS, magenta) curve is a blackbody at the median temperature of
		$84$ eV fitted by \citet{UrquhartSoria2016}, plus a $\Gamma=3$ tail
		carrying 10\% of the $0.3-10$ keV flux. As a reference, a typical
		sub-Eddington black hole binary (BHB) spectrum with a $kT_{\rm in}=1$ keV
		multi-temperature blackbody component and a power-law corona component
		carrying half of the $2-20$ keV flux is shown as the gray dotted line. The
		vertical dashed line marks the energy where that corona overtakes that
		disk. All curves in both panels are available as machine-readable tables in
		the Supplemental Material.}
	\label{fig:sed_cartoon}
\end{figure}

\begin{marginnote}[]
\entry{Structure function (SF)}{The SF measures the root-mean-square (rms) magnitude difference in a light-curve as a function of time lag $\Delta t$. Its long-lag asymptote $\SFinf$ quantifies the total variability amplitude.}
\end{marginnote}

{\it Variability in UV/Optical bands}: AGNs also show intrinsic variability in the rest-frame UV/optical, typically with smaller fractional variability amplitudes ($\sim 3-15\%$) than in the X-ray \citep[e.g.,][]{Ulrich+1997}. The characteristic variability timescale is months to years instead of days as in the X-ray, as these photons are typically produced further from the central black hole. The UV/optical variability is not always correlated with the X-ray variability \citep{Laurenti+2020}. PHL 1092 \citep{Miniutti+2009} shows a factor of 200 drop in X-ray flux over a five-year time scale, while the UV flux remains stable. 

Two complementary statistical descriptions are commonly used to quantify the variability properties -- the structure function (SF) \citep[e.g.,][]{Simonetti+1985} and power spectral density (PSD). Generally, both descriptions are well-captured by the damped random walk (DRW), a first-order continuous auto-regressive process that is used to model AGN optical light curves
\citep[e.g.,][]{Kelly+2009,MacLeod+2010}. The corresponding PSD is typically a broken power law where the breaking frequency corresponds to a damping time scale $\tau_{\rm DRW}$. Using long-baseline light curves that adequately resolve $\tau_{\rm DRW}$,
\citet{Burke+2021} established a tight, roughly linear relation between the
optical damping timescale and black hole mass over $\sim$6 orders of magnitude in $\mbh$, from local AGN down to accreting stellar-mass compact objects. It is thought that $\tau_{\rm DRW}$ is closely related to the thermal time-scale in the UV/Optical emission region. 

Multi-wavelength analysis of a large sample of quasar lightcurves in stripe 82 and Pan-STARRS 
\citep[e.g.,][]{Burke+2021} show that the excess variance and PSD amplitude are strongly anti-correlated with bolometric luminosity and Eddington ratio, while the dependence on black hole mass is weak and inconsistent across studies. All these studies also find that the variability amplitude anti-correlates with the rest frame wavelength. The blue bands typically vary more with a shorter characteristic timescale. This all suggests that super-Eddington sources will have smaller variability amplitude in the UV/Optical bands, which is in contrast to the trend seen in X-rays, e.g.\ for NLS1 galaxies.

\subsubsection{Radio Properties}
\label{sec:radio}

During the state transition of X-ray binaries, a persistent and compact relativistic jet is typically found when the source is in the low hard state, where the accretion rate is well below the Eddington value and the spectral shape is dominated by non-thermal emission. In the high/soft state, the accretion rate increases and the spectral shape becomes soft as a radiatively efficient thin disk is formed, the jet is usually quenched \citep[see the reviews][and references therein]{RemillardMcClintock2006,Done+2007}. The core radio emission falls by at least two to three orders of magnitude \citep{Fender+1999,Russell+2011}. Soft-state radio detections are correspondingly rare, and those that occur are generally attributed to the decaying remnants of ejecta launched during the transition rather than to an ongoing compact jet \citep{FenderBelloni2004}.

A similar trend also exists in AGNs. In general, radio loudness (that is the fraction of the bolometric luminosity that emerges in the radio) increases at lower Eddington ratio \citep[e.g.,][]{Sikora+2007,Ho:2008}. A natural extension of this trend is that sources suspected to be Eddington or super-Eddington tend to be radio quiet \citep[e.g.,][]{Kellermann+1989}. Interestingly, \cite{Laor+2019} also find that for radio-quiet sources, the radio slope is flat for the low Eddington sources, but steep for the high Eddington sources. Furthermore, radio-loud and radio-quiet sources display different trends between $\mbh$ and radio slope, strongly implying different origins for the radio emission. \cite{Laor+2019} thus demonstrate that the radio properties also participate in Eigenvector 1. Indeed, some of the sources we most associate with large-scale winds, like the ERQs \citep{Hwang+2018}, narrow-line Seyfert 1 galaxies \citep{Moran+2000}, and even red quasars more generally show steep radio spectral slopes, particularly when compact cores dominated by accretion are considered \citep[][]{Fawcett+2025}.


There are two possible origins for this optically thin emission that are discussed in the literature \citep[e.g.,][]{Laor+2019}. One is that the optically thin radio emission associated with radio quiet sources at high Eddington ratio is plausibly produced in a $\sim$pc-scale wind \citep[e.g.,][]{Jiang+2010,ZakamskaGreene2014},
which fits into our emerging picture that sources at high Eddington ratio often show clear signs of a wind. The other is that this optically thin radio emission emerges from the corona \citep{LaorBehar:2008}, also supported by the relatively constant ratio of $L_R/L_X$ seen in sources with high accretion rate \citep[e.g.,][]{PaulPlotkin:2026}. \citet{Chen:2024} present VLBA radio observations and show that the compactness and spectral slopes lean towards winds in the high Eddington sources.

\section{Samples of Super-Eddington AGNs and subclasses}
\label{sec:sub_classes}

In this section, we provide a summary of named sub-classes of objects that are thought likely to be super-Eddington sources. These objects span a wide redshift range from local to cosmic dawn, as well as a very wide range in luminosity. Their super-Eddington nature is typically inferred based on single-epoch black hole masses combined with a luminosity measurement. It is worth noting that the available diagnostics are heavily dependent on the redshift and selection method (e.g., \emph{WISE}-selected sources are optically weak, so we cannot compare their intrinsic optical properties with UV-selected sources). 

The redshift-luminosity distributions of the extragalactic super-Eddington sources are shown in Figure \ref{fig:redshift_lum}. We will use the properties of these samples to test the theoretical and numerical models of super-Eddington accretion.

\subsection{Ultraluminous X-ray sources (ULXs)}

Ultraluminous X-ray sources (ULXs) are point-like, off nuclear X-ray sources whose apparent isotropic luminosities exceed $10^{39}$ erg/s, which is greater than the Eddington luminosity of stellar mass black holes, $\Ledd\approx 1.25\times 10^{39} \left(\mbh/10\msun\right)$ erg/s. There have been extensive reviews of ULXs in the literature 
\citep{Roberts2007,FengSoria2011,KingLasotaMiddleton2023}.
Here we summarize the main properties of ULXs, particularly emphasizing the unique properties of these sources. While the review focuses on supermassive black holes, these ULXs provide some of the strongest evidence for super-Eddington accretion, and we will see similarities with the higher-mass black holes, which will give us important clues on what super-Eddington accretion sources may look like. 

High-quality XMM-Newton EPIC spectra of nearby ULXs are well described by two thermal-like components rather than the multi-temperature blackbody disk plus power-law that is typical for galactic X-ray binaries 
\citep{GladstoneRoberts2009,Sutton2013}. These sources are usually found to be in the ``ultraluminous state'', with a cool component at a characteristic temperature of $0.1-0.4$ keV for the soft X-ray band and a hotter component peaking at $1-3$ keV for the hard X-ray band, which typically rolls over below $\sim 10$ keV. Hard X-ray ($3-30$ keV) spectroscopy with NuSTAR also confirms that spectra of ULXs cut off well below the energies typically observed for sub-Eddington X-ray binaries \citep{Bachetti2013}. The spectra are inconsistent with the typical optically thin Comptonization model for hard X-rays, but instead require a warm, optically thick Comptonising medium. This suggests that the corona in ULXs, if it exists, likely has higher densities. ULXs near the low luminosity end $(\lessapprox 3\times 10^{39} {\rm erg/s})$ typically show a single broadened disk-like component with temperature $kT\sim 1-2.5$ keV.


There is also another distinct subclass of ultraluminous supersoft sources that do not show any detectable photons above $1.5-2$ keV 
\citep{Liu2008}. Their spectra are dominated by a blackbody-like component with typical temperature of $kT\sim 50-150$ eV and bolometric luminosity $\sim 10^{39}-10^{40}$ erg/s. The spectral shapes of these sources are similar to those of the nuclear-burning white dwarfs, except that the inferred blackbody radii and luminosity are too large to be sustained by white dwarfs. Several ultraluminous supersoft sources show clear transitions to and from the typical two-component hard spectrum of ULX \citep{Feng2016}. It has been suggested that these supersoft sources have super-Eddington accretion rates with strong outflows and they are viewed with high inclination angles \citep{FengSoria2011}. 

There is evidence of clumpy winds: blue-shifted emission or absorption features produced by Fe, Ne, O at projected velocities $\approx 0.1-0.3\ c$ \citep{Middleton+2015, Pinto+2016},
 indicating outflows in these systems. These winds are typically not static as demonstrated by time-resolved spectroscopy \citep{Pinto+2020,Middleton+2022}. ULXs typically have deeper and more strongly blue-shifted absorption when the spectra become softer. The wind kinematics in ULXs are also found to vary on timescales of hours to weeks \citep{Kosec+2021, Pinto+2021}, which suggests the existence of very small scale structures in the wind (for example a clumpy wind as discussed in \S \ref{sec:clumpy_wind}). 





These spectral classes are compared directly in the right panel of Figure~\ref{fig:sed_cartoon}, which shows the median spectral model of each class, each at its own luminosity, in the same spirit as Figure 2 of \citet{Kaaret2017}. We also add a typical spectrum for a sub-Eddington X-ray binary (in the steep powerlaw state) for reference. When the accretion rate is increased from the sub-Eddington X-ray binary, to the broadened disk and ultra-luminous state, the hard X-ray component is getting weaker and more energy is emitted in the longer wavelength band. We will see similar behavior in super-Eddington AGNs below, as they are typically X-ray weak. The supersoft curve peaks near 0.3 keV and radiates about half of its power below the observed band. These photons are thought to be produced at the photosphere of an optically thick wind with radii $\sim 10^3-10^4\ r_g$ from a stellar mass black hole.


\subsection{High Eigenvector 1 Sources}

If one sorts classic UV-bright AGN by eigenvector 1, it is thought that the sources with the highest \ion{Fe}{ii}/H$\beta$ and the weakest [\ion{O}{iii}]/H$\beta$ radiate near or even above their Eddington limit. When these sources are selected in the optical at low redshift, they are often classified as ``narrow-line'' Seyfert 1 galaxies \citep[][NLS1, see spectrum in the left panel of Fig. \ref{fig:sed_cartoon}]{OsterbrockPogge:1985}, technically defined to have FWHM$_{\rm H\beta} < 2000$~km/s. At higher redshift and luminosity, high eigenvector 1 sources are often classified as weak-lined quasars \citep[e.g.,][]{Leighly+2007,Ni+2018}. These are selected to have low equivalent width (or ``weak'') broad \ion{C}{iv} lines, often strongly blue-shifted. Indeed, declining \ion{C}{iv} equivalent width and increased blueshift are part of eigenvector 1 \citep[e.g.,][]{ShenHo2014}. The proto-typical weak-lined quasar PHL1811 \citep[][]{Leighly+2007} looks like a classic high eigenvector 1 source in the rest-frame optical as well, with strong \ion{Fe}{ii} and weak [\ion{O}{iii}], suggesting that weak-line quasars and narrow-line Seyfert 1 systems may be in a similar physical state, albeit with differing typical luminosities.

\begin{marginnote}[]
\entry{Soft X-ray excess}{
The soft X-ray excess is a smooth additional emission component below $\sim2$~keV over the extrapolation of the $2$--$10$~keV power law produced by the corona. It can carry a luminosity comparable to that of the hard X-rays for the brightest cases. If it is fitted with a blackbody, the temperature comes out at $kT\approx0.1$--$0.2$~keV and, crucially, is nearly the \emph{same} in every object and independent of black hole masses and accretion rates \citep{GierlinskiDone2004}.}
\end{marginnote}

Both types of object have distinct X-ray properties, with a prominent soft X-ray excess and rapid X-ray variability \citep{Boller+1997,Ni+2022}, and a tail to X-ray weakness \citep[e.g.,][]{Desroches+2009}. They likewise show less UV/optical variability than typical broad-line objects \citep[e.g.,][]{Klimek+2004}. Along with the idea that these sources are Eddington or super-Eddington, there is clear evidence for outflowing material on pc to kpc scales from the extensive blue wing of [\ion{O}{iii}] seen most prominently in high EV1 sources \citep[e.g.,][]{Schmidt+2018}. In fact, in some extreme cases, the entire [\ion{O}{iii}] line is observed to be dramatically blue-shifted \citep[e.g.,][]{Phillips:1976,Komossa+2008} and this phenomenon is seen to be linked with blue-shifts in the broad \ion{C}{iv} line \citep[e.g.,][]{LeighlyMoore:2004}, which is believed to originate on the scale of the broad-line region \citep[e.g.,][]{BaskinLaor:2005,Richards+2011}. Ultra-fast outflows (UFOs) are also detected in high Eddington NLS1s, which show blueshifted \ion{Fe}{xxv} or \ion{Fe}{xxvi} absorption with the implied velocity $\sim 0.05c-0.3c$. The absorption strength is anti-correlated with continuum flux and is consistent with radiative driving \citep{Pinto+2018}. 

Another subset of UV-bright AGN that might harbor a significant fraction of Eddington-limited or super-Eddington sources are the very highest-redshift UV-drop--out selected quasars, with $z>5.5$ and $L_{\rm 1450} > 5 \times 10^{45}$~erg/s. Based on scaling relations, $\sim 35\%$ are estimated to be super-Eddington \citep{Kurk+2007,Fan+2023}. There are hints that the population may skew towards higher Eddington ratios when compared with UV-selected AGN at $z<4$. It is known that at the highest UV luminosities, the optical to X-ray luminosity ratio is considerably higher such that the fraction of the bolometric luminosity released in the X-ray drops dramatically \citep[e.g.,][]{Lusso+2012}. At the highest redshifts, UV-selected AGN show similar trends between the UV and X-ray \citep[e.g.,][]{Shemmer+2006} but the spectral slope in the X-ray is much steeper and therefore softer than lower-redshift sources at a similar luminosity \citep[e.g.][]{Zappacosta+2023}. There is a hint that these AGN may also display a higher fraction of large X-ray variability events \citep{Vito+2022feedback}. The fraction of weak-lined quasars seems to be considerably higher -- from 1\% at $z<4$ to $\sim 12\%$ at $z>5.6$ \citep[][]{Shemmer+2010,Banados+2016} and the fraction of $z>5$ UV-selected AGN showing [\ion{O}{iii}] outflows with $v > 1000$~km/s is $\sim 20\%$, roughly 4 times higher than seen at Cosmic Noon \citep{Liu+2026}. The steeper hard X-ray slopes, increased incidence of weak \ion{C}{iv}, and lowered variability all point to a higher Eddington/super-Eddington population in UV-selected quasars at $z>5$, even when compared with similarly UV-luminous sources at Cosmic Noon.

A subset of broad-absorption line (BAL) quasars, FeLoBALs, show dramatic low-ionization BALs and even \ion{Fe}{ii} absorption. While BAL quasars are quite common, comprising roughly 10-20\% of quasars \citep[e.g.,][]{HewettFoltz:2003}, the LoBALs that show MgII absorbers are rare. Only a subset of these also show absorbers in \ion{Fe}{ii} and sometimes other Fe-peak elements \citep[e.g.,][]{deKool+2002}. It is thought that BALs are present in most systems, but only seen at a narrow range of orientation \citep[e.g.,][]{Gallagher+2006}. In addition, there are at least a subset of the FeLoBALs that may trace a super-Eddington phase \citep[][]{Choi+2022,Leighly:2024}.

\subsection{Extreme Red quasars (ERQs) and Hot dust-obscured galaxies (HotDOGS)}

While classic luminous accreting black holes are selected in the UV, with the advent of \emph{Spitzer} and \emph{WISE}, near-to-mid infrared selection also became possible. The Extremely Red Quasars \citep[ERQs;][]{Hamann+2017} and Hot Dust-Obscured Galaxies (HotDOGS) \citep[e.g.,][]{Dey+2008,Tsai+2015} are selected to be extremely red and found to be very rare. By selection, both populations are red in both mid-infrared and observed-frame optical to mid-infrared color. ERQs are defined to have $i-W3 > 4.6$, while HotDOGs are selected as W1W2 drop-outs. In both cases, the sources are extremely red in the rest-frame near-to-mid--infrared and very luminous in the mid-infrared. The redness is typically interpreted as arising from substantial reddening on small scales, although there is still debate about the origin of the observed UV, and thus the relative roles of partial covering, dust reddening, and scattered light \citep[e.g.,][]{Assef+2016,Alexandroff+2018}.

The sources are typically found at Cosmic Noon ($2<z<4$), and because of the sensitivity of WISE, only the most luminous (rare) sources are detectable at these distances. They have implied bolometric luminosities $L_{\rm bol} \approx 10^{46}-10^{48}$~erg/s, with the HotDOGs residing at the higher end of that range. The ERQs by selection have SDSS spectra in the rest-frame UV, where they display boxy \ion{C}{iv} profiles with high rest-frame equivalent widths \citep{Hamann+2017}, and blue rest-UV color. The HotDOGs show similar distributions in rest-UV line properties \citep{Jun+2020}. Polarization studies suggest that the UV emission is dominated by scattered light, at least in some cases \citep[][]{Zakamska+2023,Assef+2026}.

HotDOGs and ERQs are faint in the X-ray relative to their mid-infrared luminosities \citep[e.g.,][]{Goulding+2018}. In part, this faintness can be ascribed to gas absorption, which might be expected given the very red color of the optical--to-mid-infrared spectra. However, there is some evidence that the sources are also intrinsically X-ray weak \citep[e.g.,][]{Ma+2024}. In the rest-frame optical, some targets show clear evidence of broad Balmer lines, while others do not \citep[e.g.,][]{Jun+2020,Ma+2024}. However, an appreciable fraction show dramatic evidence of large-scale outflow in the form of [\ion{O}{iii}] lines with FWHM$=1000-4000$~km/s \citep{Zakamska+2016,Jun+2020}. In the rest-frame UV, there is some evidence of outflow in the polarization properties of the \ion{C}{iv} line, which shows a dramatic 90 deg shift in angle across the line \citep{Alexandroff+2018}, which may suggest that the \ion{C}{iv} emerges from a wind on the scale of the torus \citep{Zakamska+2023}.

\subsection{Little Red Dots}

One of the most exciting early discoveries from the \emph{James Webb Space Telescope} has been a new class of object called `Little Red Dots' \citep[LRDs;][]{Matthee+2024}, characterized by compact sizes -- upper limits of 10s of pc based on lensing \citep[e.g.,][]{Furtak+2024}, a `v-shaped' spectral energy distribution that is blue in the rest-frame UV and red in the rest-frame optical \citep[e.g.,][]{Labbe+2023,Kocevski+2023}, with the inflection point of the `v' typically occurring at or redward of the $n=2$ Balmer limit \citep{Setton+2025vshape}, and broad Balmer lines \citep[e.g.,][]{Greene+2024}. Leveraging the well-defined selection function of the RUBIES program \citep{degraaff+2025}, \citet{Hviding+2026} find that if one selects on any two of these three properties, at least 80\% of the sample displays the third property, strongly suggesting that these properties together do define a class of objects. In addition to these characteristics that we use to select sources, the LRDs are also typified by prominent Balmer and He absorbers \citep[e.g.,][]{Matthee+2024,Lin+2024,Juodzbalis+2026}.

The power source of the Little Red Dots remains a topic of debate. The strong Balmer break and red optical color were suggestive of massive galaxies \citep{Labbe+2023,Baggen+2024}, while the compact size and broad lines pointed towards accreting black holes \citep{Kocevski+2023,Harikane+2023}. The accreting black hole hypothesis is challenged by the highly irregular broad-band SED, which is characterized by extremely weak X-ray emission \citep[e.g.,][]{Yue+2024} and lack of a clear hot dust signature out to $\sim 5 \, \mu$m \citep[e.g.,][]{Williams+2024}, although some hot dust excess is seen in local analogs \citep[][]{Lin+2026Egg,Ji+2026lord}. Furthermore, there is little sub-millimeter emission, casting doubt on the possibility that the sources look red because of dust absorption \citep[e.g.,][]{Setton+2025,Casey+2025}. There is no evidence for variability on timescales $\lessapprox 5$~yr rest-frame, despite significant effort to detect it \citep[e.g.,][]{Liu+2026twinkle,Stone+2026}. Some LRD optical spectra have continuum shapes that cannot be modeled with any combination of any stars that we know of \citep[e.g.,][]{Ma+2025404,degraaff+2025cliff,Naidu+2026}. This fact, combined with their compactness and the prevalence of very broad Balmer lines, has led to a growing consensus that they are powered by accretion onto massive black holes. However, many other alternatives are still serious contenders, such as supermassive stars \citep[e.g.,][]{Nandal+2026,Chisholm+2026} or tidal disruption events \citep[e.g.,][]{Bellovary:2025}.

As it happens, many of the most perplexing properties of the LRDs are in common with other super-Eddington sources discussed in this review. Specifically, X-ray weakness and low variability are often associated with high accretion rate objects. This has led to the suggestion that the overall phenomenology of LRDs may be explained by super-Eddington accretion onto a massive central black hole \citep{Lambrides+2026,Liu+2025superedd}. We will revisit some of the modeling progress on LRDs in \S \ref{sec:lrdmodel}.

\begin{marginnote}[]
	\entry{Changing-Look AGNs}{Changing look AGNs are a class of AGN showing dramatic changes in luminosity and spectra over time scales of a few months to years \citep{RicciTrakhtenbrot2023}. They are divided into two sub-classes based on the cause of the change. The ``changing-obscuration objects'' show strong variability in the line-of-sight column density, so the apparent change is caused by changing gas obscuring the central engine, which itself does not change. ``Changing-state AGNs'' have continuum emission and broad emission lines that can appear and disappear. The accretion disks in these objects change significantly, which produces different spectra of the continuum emission. The broad lines then respond to the changes in continuum. }
\end{marginnote}

	

\subsection{Transients}
Transients provide unique opportunities to probe how the properties of the accretion disks change when the mass accretion rate varies across the Eddington limit on a timescale that is much shorter than what a steady state accretion disk would predict. As an example, \emph{1ES 1927+654} was originally identified as a Type-2 AGN with no broad emission lines, and was observed in this state for many years. Then its optical and UV luminosity was found to increase by a factor of $\sim 20$ in 2018, marking the beginning of a transient event. It reached a peak value $\sim 2\times 10^{44}$ erg/s, which then slowly declined \citep{Trakhtenbrot+2019}. Broad emission lines also started to show up a few weeks after the beginning of the event. At the same time, the X-ray luminosity dropped from $10^{43}$ erg/s to $\approx 10^{40}$ erg/s in the first 200 days. However, the subsequent year the X-ray luminosity reversed and reached $\approx 2\times 10^{44}$ erg/s over the course of a year \citep{Ricci+2020}. During the high state, the X-ray spectrum has a broad peak in the $0.3-2$ keV range and drops quickly above $2$ keV with photon index $\Gamma \approx 3$. This is also quite different from the much harder X-ray spectrum of this source before the outburst. \cite{Ricci+2020} suggests that this outburst is caused by the disruption of a star near the black hole. The spectral evolution of this source can be understood based on general trends observed in AGN spectra as a function of mass accretion rate. During the initial outburst, a lot of gas is likely dumped to the inner region of the disk as suggested by \cite{Ricci+2020}, which causes super-Eddington mass accretion onto the black hole. At this point, the inner region becomes optically thick. This will cause the photons to be thermalized and the X-ray luminosity to drop, as found in many super-Eddington AGNs. The accretion rate drops after a few hundred days and eventually a disk is formed with strong soft X-ray emission similar to NLS1s. 
It should be noted that although classical theoretical models for TDEs predict that the mass fall back rate should be much larger than the Eddington value, the observed luminosity of most TDE candidates is always below the Eddington value and they do not show the same temporal evolution of X-rays as in \emph{1ES 1927+654} 
\citep{Lodato+2011,Yao+2023}.



\begin{figure}[htbp]
	\centering
	\includegraphics[width=\linewidth]{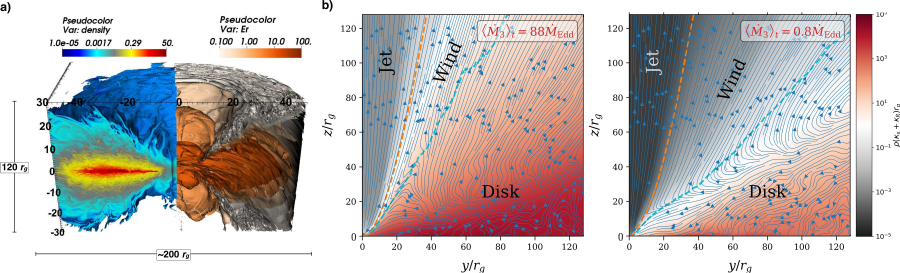}
	\caption{Typical structures of super-Eddington accretion flows onto stellar
	mass black holes from radiation MHD simulations.
	(\emph{a}) A snapshot of the three-dimensional radiation MHD simulation of
	\cite{Jiang+2014a}, with a disk accreting at $\approx 22\,\Medd$ onto a
	$6.62\ \msun$ black hole. The two halves show volume renderings of the same
	snapshot: gas density on the left and radiation energy density $E_r$ on the
	right, both in code units and on logarithmic scales. Axis ticks are the
	cylindrical radius and height in units of $2\ r_g$, and the scale bars mark
	the $\approx 200\ r_g \times 120\ r_g$ region shown.
	(\emph{b}) Time- and azimuthally-averaged structures from the radiation
	GRMHD simulations of \cite{Zhang+2025}, in the poloidal plane with the
	cylindrical radius $y$ and height $z$ in units of $r_g$, for a
	super-Eddington run with $\langle\dot{M}\rangle_t = 88\,\Medd$ (left) and a
	sub-Eddington run with $\langle\dot{M}\rangle_t = 0.8\,\Medd$ (right), both
	within the inner $128\ r_g$. Colors show the total (scattering plus
	Rosseland mean absorption) optical depth per gravitational radius,
	$\rho(\kappa_s+\kappa_R) r_g$, and blue lines with arrows are streamlines of
	the poloidal velocity. The orange and cyan dashed lines separate the three
	dynamically distinct regions labeled in each panel: the magnetically
	dominated, evacuated jet funnel along the rotation axis, the unbound wind at
	intermediate polar angles, and the bound, turbulent accretion disk near the
	midplane.}
	\label{fig:simulation_XRB}
\end{figure}

\section{Theoretical Models for Super-Eddington Accretion}
\label{sec:simulation}
The standard thin disk model assumes that energy generated via the accretion process is radiated away locally. When the mass accretion rate gets close to or exceeds the Eddington value, the surface density will increase and photons may not have time to escape before they are advected towards the black hole. At the same time, when the total luminosity reaches the Eddington limit, outflow from the disk is unavoidable.

\subsection{Semi-analytic Models}
\label{sec:model}
Models for super-Eddington accretion disks have been reviewed extensively in recent years \citep{AbramowiczFragile2013,Inayoshi+2020,JiangDai2025}. We provide a brief summary of these semi-analytical models here, which will be useful for the remaining discussion and comparisons with observations.

Most semi-analytic super-Eddington accretion disk models make the following assumptions, which have important implications for their predictions. 1) The flow is assumed to be steady state and axisymmetric. Time-dependent properties (such as limit-cycle behavior caused by instabilities, variability) are perturbations from the steady state disk models. 2) The equations of mass, momentum and energy conservation are integrated over height, reducing the problem to a set of coupled ordinary differential equations as a function of radius. The vertical structure of the disk is constructed based on photon diffusion and an assumed vertical dissipation profile. 3) Angular momentum transport is parameterized by the famous $\alpha$ ansatz proposed by \cite{ShakuraSunyaev1973}, where the vertically integrated stress is assumed to be proportional to the total pressure. This is typically the thermal pressure (gas and radiation pressure) but can also include magnetic pressure in some models. 

The earliest super-Eddington models were the Polish doughnuts, which were steady state, radiation pressure supported rotating tori \citep{Kozlowski+1978,Abramowicz+1980}. These tori radiate locally at the Eddington flux on the surface and the total luminosity $L$ scales with the accretion rate $\dot{M}$ as $L/\Ledd \approx 1 + \ln \left(\dot{M}/\Medd\right)$. The thickness of these tori is comparable to the radius with a thin funnel around the rotating axis. Many properties of the Polish doughnuts are similar to the slim-disk model \citep{Abramowicz+1988}, which improved the simplified model to include more self-consistent energy transport within the disk. 

\begin{marginnote}[]
\entry{Funnel}{This is the region near the rotation axis of the disk, typically considered a cone with a few degrees of opening angle.}
\end{marginnote}

The slim-disk model is an extension of the standard thin disk model \citep{ShakuraSunyaev1973} to the super-Eddington regime that includes the following modifications: 1) local energy transport is not only contributed by \emph{vertical} photon diffusion, but also \emph{radial} advection. This is because the vertical photon diffusion speed is $\sim c/\tau$, where $\tau$ is the vertical optical depth from the disk midplane to the surface, while the radial advection is controlled by the inflow speed $v_r$. Super-Eddington disks are expected to have smaller $c/\tau$ and larger $v_r$ compared to sub-Eddington disks. Therefore, radial advection should be more important in these disks than in thin disks. This is also the key concept of the photon trapping radius $R_{\rm tr}$ \citep{Begelman1978}, which is the radius where photons are advected towards the black hole instead of diffusing out in spherical geometry
\begin{eqnarray}
	R_{\rm tr}=\frac{\dot{M}c^2}{\Ledd}r_g. 
\end{eqnarray}
2) The gravity from the central black hole is balanced by a combination of radial pressure gradient and rotation. Therefore, the angular velocity of the disk is generally sub-Keplerian. When the accretion rate increases, the slim disk model does not approach the spherical limit. Due to significant radial advection, the density weighted scale height will saturate at $70\%-80\%$ of the radius no matter how large the accretion rate is \citep{Lasota+2016}.

It has been recognized that large radiation pressure will expel the excess accreting gas when the disk becomes locally super-Eddington, which was already pointed out in the thin disk model \citep{ShakuraSunyaev1973}. \cite{Poutanen+2007} modified the slim disk model by including the mass loss by the wind for the region $r<R_{\rm tr}$. If the local radiation flux is assumed to be the Eddington value inside $R_{\rm tr}$, then the mass accretion rate will decline inward as $\dot{m}\propto R$. Outside the trapping radius, photon diffusion is the dominant cooling process and the solution is reduced to the standard thin disk.\footnote{This can be also understood by comparing the viscously dissipated radiation flux emitted from the face of the disk $F_{\rm disk}$ and the Eddington flux $F_{\rm Edd}$. For a disk with a constant mass accretion rate, $F_{\rm disk}\propto R^{-3}$ while the Eddington flux $F_{\rm Edd}\propto R^{-2}$. Therefore, $F_{\rm disk}$ will always drop below $F_{\rm Edd}$ when $r > R_{\rm tr}$.} Because the wind launched from the inner region is optically thick when the mass accretion rate is larger than $\approx 2.5 \Medd$, the radiation emitted from the disk is reprocessed into a soft blackbody with typical effective temperature smaller than the typical value of the disk. 

All the models mentioned above only consider gas and radiation pressure to determine the accretion disk structure. Following the well-known $\alpha$ ansatz proposed by \cite{ShakuraSunyaev1973}, angular momentum transport is assumed to be carried out by a torque that is proportional to the total pressure. Since the magneto-rotational instability \citep{BalbusHawley1991,BalbusHawley1998} is believed to be the physical mechanism that generates the turbulence for angular momentum transport in disks, \cite{BegelmanPringle2007} proposed that the magnetic field can be amplified by MRI to a state in which magnetic pressure becomes larger than thermal pressure. It is assumed that the Alfv\'en speed of the saturated magnetic field is the geometric mean of the orbital and sound speeds \citep{PessahPsaltis2005}. The vertical structure of these magnetically elevated disks are refined and improved by considering the balance between the magnetic field amplification via the dynamo process and the escape of magnetic fields due to buoyancy. This configuration is subject to the Parker instability \citep{Begelman+2015}. One consequence of the unstable Parker mode is the enhanced vertical energy transport due to the advective radiation energy flux associated with magnetic buoyancy \citep{Jiang+2014a}. Magnetically elevated disks have the same radial effective temperature dependence as in the thin disk model, but at a significantly larger disk scale height and smaller surface density compared with a thin disk model at the same mass accretion rate. The inflow speed in these disks is typically much larger than the inflow velocity at the same radius in the thin disks due to a large disk scale height, which can explain the rapid change of accretion rate in AGNs over the time scales of a few years \citep{DexterBegelman2019}. 

\cite{Hopkins+2024a,Hopkins+2024b} proposed an even more extreme magnetic-pressure--dominated model for super-Eddington accretion disks. In this model, the turbulent toroidal magnetic fields in the disk are not amplified by MRI in situ. Instead, they are carried inward by the accretion flow via a flux-freezing condition from the galaxy scale. Therefore, the magnetic field amplitude is related to the gas density as $B\propto \rho^{2/3}$. Magnetic pressure can be $\sim 10^2-10^4$ times the gas pressure in this model. The flow is assumed to have supersonic but trans-Alfv\'enic turbulence with the turbulent velocity varying with radius as $R^{-1/3}$ while the inflow speed is roughly $\propto R^{-7/6}$. In the standard slim disk model, the effective inflow speed is proportional to the thermal sound speed $c_s$. For the magnetically elevated disk model discussed in the previous paragraph, the effective inflow speed is $\propto c_s/\beta^{0.5}$ with magnetization parameter $\beta\sim 0.1$. As $\beta$ is smaller by another factor of $\sim 100$ in this model, the effective inflow speed is larger by another order of magnitude. The surface density of the disk in this model is smaller than the slim disk model by a factor of $\sim 10^3-10^4$ because of this large inflow speed as constrained by the continuity equation. This low density causes the electron scattering optical depth to drop below unity beyond $10^4$ gravitational radii at the Eddington accretion rate. This transition region will move closer to the black hole when the accretion rate drops.

In summary, all semi-analytical models for super-Eddington accretion point out that the disk will be thicker than sub-Eddington disks and that radial advection is important for energy transport. Outflow is also an unavoidable outcome of these disks. However, these models need to make different assumptions about magnetic fields, which can result in quite different disk structures. It is also hard for semi-analytical models to make quantitative predictions about the properties of the outflows and how they vary with accretion rate and distance from the black hole. All these questions need to be answered by self-consistent radiation MHD simulations.

\begin{figure}[htbp]
	\centering
	\includegraphics[width=0.9\linewidth]{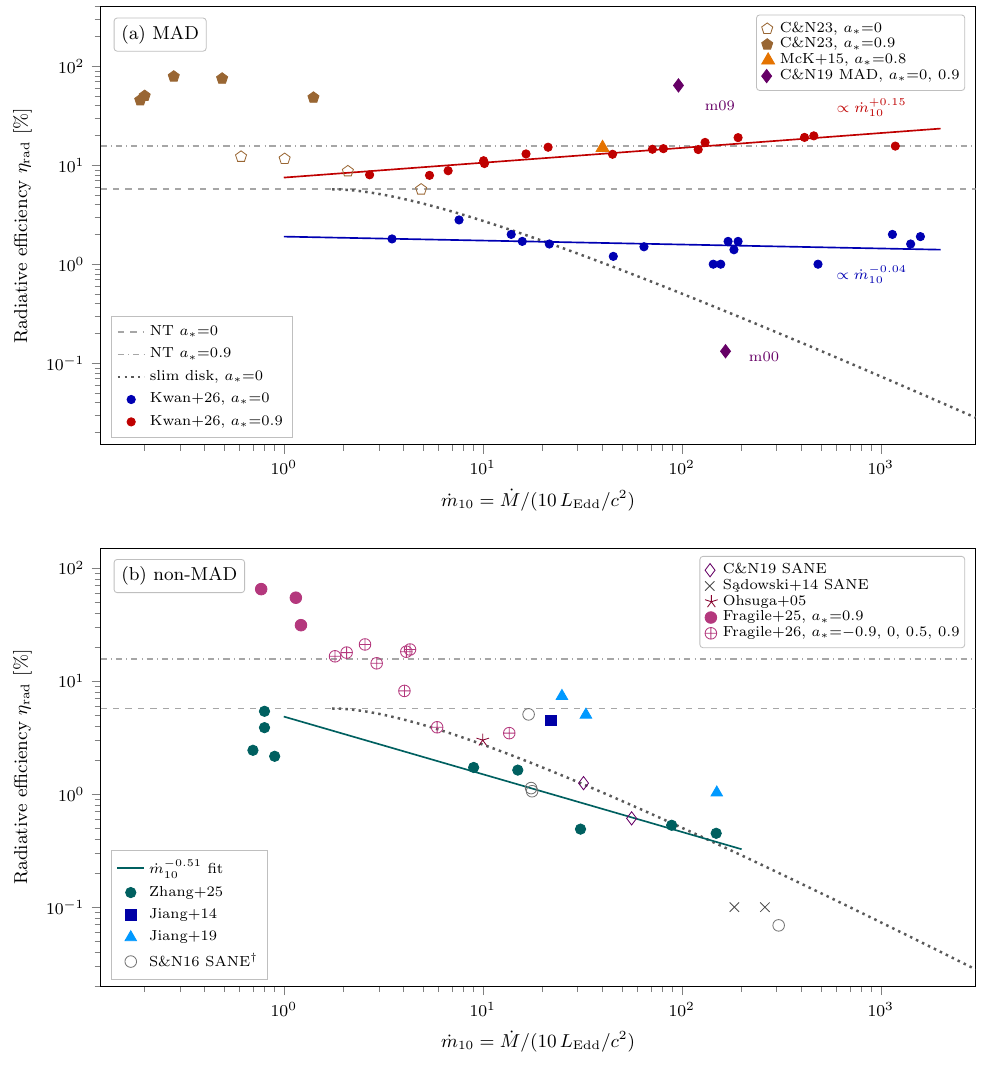}
	\caption{Radiative efficiency $\eta_{\rm rad}\equiv L_{\rm rad}/(\dot{M}c^2)$
	as reported by numerical simulations of super-Eddington accretion. All the mass accretion rates 
	in the published simulations have been converted to the same definition of Eddington mass accretion rate $10\Ledd/c^2$. Grey lines mark the Novikov--Thorne thin-disk values
	\citep{NovikovThorne1973}: $5.7\%$ for $a=0$ (dashed) and $15.6\%$ for
	$a=0.9$ (dash-dotted). The dotted curve is the slim-disk expectation
	\citep{Abramowicz+1988} as $L\simeq\Ledd\left[1+\ln(\dot{M}/\dot{M}_{\rm crit})\right]$. 
	The top panel (a) is for the MAD runs adopted from \cite{McKinney+2015,CurdNarayan2019,CurdNarayan2023,Kwan+2026} while the bottom panel (b) is for the non-MAD runs adopted from \cite{Ohsuga+2005,Jiang+2014a,Jiang+2019,Sadowski+2014,Sadowski+2016b,CurdNarayan2019,Zhang+2025,Fragile+2025,Fragile+2026}. Notice that the efficiencies of
	\cite{Sadowski+2016b} are not quoted by those authors but derived here from
	their tabulated radiative luminosities. The MAD runs typically have a much higher radiative efficiency than the non-MAD runs while the non-MAD runs show a similar trend of declining radiative efficiency with increasing mass accretion rate. Every plotted point, with the accretion-rate conversion applied to it, is available as a machine-readable table in the Supplemental Material.}
	\label{fig:efficiency}
\end{figure}

\begin{figure}[htbp]
	\centering
	\includegraphics[width=0.85\linewidth]{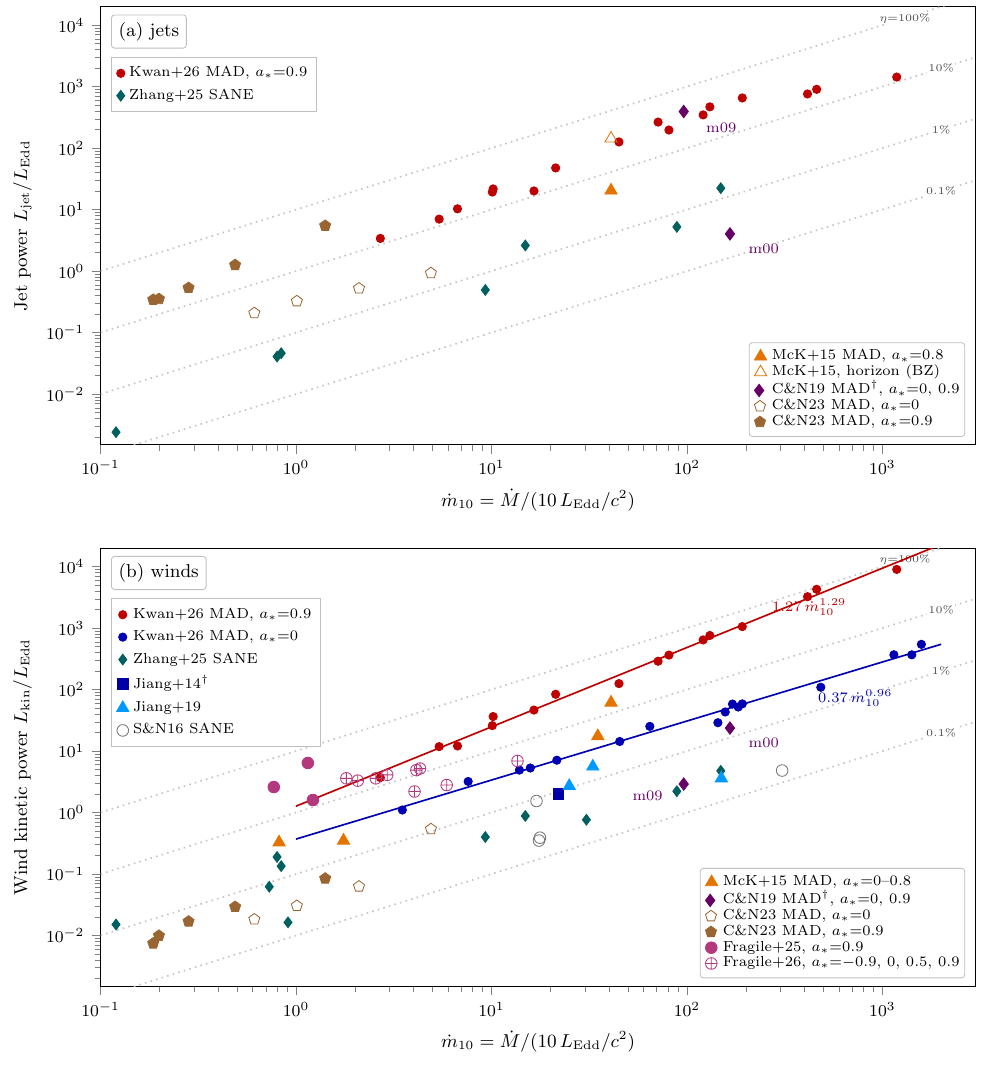}
	\caption{Mechanical power carried by the jet (panel a) and by the wind
	(panel b) in numerical simulations of super-Eddington accretion, in units
	of $\Ledd$. Accretion rates are converted to
	$\dot{m}_{10}\equiv\dot{M}/\left(10\Ledd/c^2\right)$ exactly as in Figure
	\ref{fig:efficiency}. The grey dotted diagonals are
	lines of constant efficiency.
	Each group roughly splits the narrow, strongly magnetized and mildly relativistic polar
	outflow (jet) from the slower, mass-loaded material around it (wind). 
	The two symbols from 
	\cite{McKinney+2015} in panel (a) give the jet power at $400\ r_g$
	(filled) and the Blandford--Znajek power on the horizon (open), a factor
	of seven larger because the wind absorbs most of the jet energy in
	between. The solid lines in panel (b) are the fits
	$L_{\rm kin}/\Ledd=0.37\dot{m}_{10}^{0.96}$ and
	$1.27\dot{m}_{10}^{1.29}$ of \cite{Kwan+2026} for $a=0$ and $0.9$. Values marked
	$\dagger$ from \cite{Fragile+2026} are derived here from tabulated quantities rather than quoted. Every plotted point is available as a machine-readable table in the Supplemental Material.}
	\label{fig:kin_efficiency}
\end{figure}

\subsection{Models for the Spectra of Super-Eddington Accretion Disks}
\label{sec:spec_model}
The semi-analytical slim accretion disk models generally predict that the disk aspect ratio, the ratio between disk scale height and radius, is a constant value $\approx0.45$ and the effective temperature follows the scaling $T_{\rm eff}\propto \Ledd^{1/4}r^{-1/2}$ \citep{Wang+1999}. 
Even for models including potential mass loss from the disk so that the mass inflow rate decreases towards the black hole, as long as the wind is optically thin, the disk effective temperature still follows the same scaling \citep{Watarai_Fukue_1999,Kitabatake+2002_Wind_spec}. The maximum blackbody temperature scales with the black hole mass as $T_{\rm eff, max}\simeq 0.2\left(\mbh/10^5\msun\right)^{-1/4} {\rm keV}$ \citep{Mineshige+2000} and the size of the emitting region with the maximum energy decreases 
from $\approx 20 r_g$ for the sub-Eddington case to $\approx r_g$ for the super-Eddington case in this model. The predicted spectra of Super-Eddington accretion disks are superposition of multi-temperature black body spectra based on the effective temperature of the disk, which peaks around $\sim 0.1~{\rm keV}$ to $10~{\rm keV}$ when the mass accretion rate varies from $\Ledd/c^2$ to $\sim 1000 \Ledd/c^2$ \citep{Mineshige+2000} for a $10^5\msun$ black hole.

The predicted spectra are significantly modified when an optically thick outflow driven from the accretion disk is considered. \cite{Poutanen+2007} assume that an optically thick wind will be launched within the spherization radius $r_{\rm sp}$, which is defined as the radius where the locally dissipated flux reaches the local Eddington flux, so that the luminosity released inside $r_{\rm sp}$ is $\approx\Ledd$. It also follows the same scaling as the trapping radius $R_{\rm tr}$, with $r_{\rm sp}\approx0.8\,R_{\rm tr}$ in their model.
The effective temperature at the spherization radius is $T_{\rm sp}\approx \left(M/10\msun\right)^{-1/4}\left(\dot{M}/\Medd\right)^{-1/2}$ keV, which is typically lower than the temperature of the underlying disk. The temperature at $r_{\rm sp}$ decreases with increasing mass accretion rate as the photosphere radius increases. The predicted spectra from this model are also viewing angle dependent. Face-on observers will see the inner region of the disk with high energy photons while edge-on observers are only able to see photons produced at $r_{\rm sp}$, which will be soft X-rays for stellar mass black holes and UV/Optical for supermassive black holes. These models with optically thick outflows are more broadly consistent with the observational properties of ULXs. Similar outflow models have also been applied to understand the properties of Little Red Dots \citep{LiuFeng+2026}.

A physically motivated empirical model fitted to narrow-line Seyfert 1 galaxies has also been proposed for the spectra of Super-Eddington accretion disks \citep{KubotaDone2019}. This model assumes three different regions in the disk in order to reproduce the observed broad band spectra of narrow-line Seyfert 1s. It includes an inner hot Comptonizing region and an intermediate warm Comptonizing region to produce the soft X-ray excess, with a slim disk model assumed to extend across the whole radial range. The resulting spectrum is the same as the predictions from the standard slim disk model for the low energy photons $(< 0.1 {\rm keV})$. The model for X-rays is adjusted to reproduce the spectrum slope of NLS1, which may not be applicable for other super-Eddington sources. 

One characteristic property of the magnetic-pressure--dominated disks described in section \ref{sec:model} is the low surface density. This has direct consequences for the emission spectra of these disks. Scattering typically dominates over absorption in the atmosphere, which will make harder, not thermalized spectra. Any wind generated from these disks may not be optically thick. 

\subsection{Numerical Simulations of Super-Eddington Accretion Disks}
Numerical simulations typically evolve the ideal magneto-hydrodynamical equations, coupled with radiation transport equations, which can solve the thermodynamics of the gas and track the photons self-consistently. For given initial and boundary conditions, simulations can evolve the structures of accretion disks without the assumptions typically made for semi-analytic models. 
Some key questions that can be potentially answered by these simulations include:
\begin{itemize}
	\item What is the radiative efficiency of these disks as a function of mass accretion rate
	\item What is the angular distribution of the accretion flow that will determine the viewing angle effects of these objects
	\item How much mass is carried away by the wind and the kinetic energy flux associated with the wind
	\item What is the predicted spectrum of the disk as a function of mass accretion rate
	\item What are the predictions of variability from the simulations
	\item What is the role of magnetic fields in super-Eddington accretion disks.
\end{itemize}

\subsubsection{Numerical Simulations of the Inner Region}
\label{sec:sim_inner}
Most numerical simulations of super-Eddington accretion disks focus on the inner region of the disk ($r\lesssim 100 r_g$) for two reasons. First, this is the region where most of the accretion power is produced for an accretion disk with a constant mass accretion rate. Second, numerical simulations are limited by the affordable spatial resolution to resolve the characteristic structures of the disks (such as the disk scale height), as well as the number of integration steps to achieve a characteristic timescale (such as inflow time). The smallest timescales and spatial scales are also dominated by the inner region. Therefore, only this part of the disk can be studied in great detail with all of the relevant physical processes included. 

The first two-dimensional radiation hydrodynamic simulations of super-Eddington
disks~\citep{Eggum+1988,Okuda2002,Ohsuga+2005} modeled angular momentum transport
with the $\alpha$ prescription and treated the radiation--gas coupling and photon
transport with the flux-limited diffusion (FLD) approximation. These
calculations confirmed the basic picture of the semi-analytical models: super-Eddington accretion is possible
when mass flows in near the midplane while photons escape through the funnel
region along the rotation axis. The radial temperature gradient is small and the midplane optical depth
is large, so radiation force does not stop the inflow there. The effect of photon trapping in the super-Eddington regime is basically confirmed in these simulations, where the radiative luminosity is only mildly above the Eddington value with logarithmic dependence on the accretion rate. These simulations also show that a significant optically thick
outflow is launched along the polar direction, although they 
generally were not run long enough to determine a reliable outflow rate.

Two developments were recognized as essential. First, since magneto-rotational instability (MRI) is believed to generate the turbulence
responsible for angular momentum transport~\citep{BalbusHawley1991,BalbusHawley1998},
magnetic fields must be included, and fully three-dimensional simulations are
needed to capture turbulence correctly and allow magnetic fields to produce
extra buoyancy for vertical energy transport \citep{Jiang+2014a} or photon bubbles~\citep{Begelman2002}. Second, radiation transport must be modeled
accurately, because radiation pressure greatly exceeds gas pressure---especially
for supermassive black holes---and because anisotropic transport near the funnel governs the outflow launching and the viewing angle dependence of the observables. For supermassive black holes, the typical gas temperature is in the range of $\sim 10^4-10^6$ K. This is a very generic result when the Eddington luminosity is emitted via the blackbody thermal emission distributed over an area of $4\pi r^2$ where the radius $r$ scales with the black hole's gravitational radius. 
Opacity sources contributed by metals can be more important than electron scattering in these disks and they need to be captured to get the coupling between the gas and photons correctly \citep{Jiang+2016,JiangBlaes2020}.

\begin{marginnote}[]
\entry{M1 closure}{M1 closure is a method to evolve the radiation transport equation by approximating the radiation field as a fluid. It only evolves the radiation energy density $E_r$, flux $\mathbf{F}_r$ and closes the system by an analytic expression using $E_r$ and $\mathbf{F}_r$.}
\end{marginnote}

The general properties of super-Eddington disks have been studied with two- and
three-dimensional general relativistic radiation MHD (GRMHD) simulations using
the M1 closure \citep{McKinney+2014,Sadowski+2014,McKinney+2015,Sadowski+2015,Utsumi+2022}.
These simulations typically start from a rotating torus at $15$--$30\ r_g$ with a single poloidal field loop seeding the MRI; when enough magnetic flux accumulates near the horizon, the flow enters the magnetically arrested disk (MAD) state. 
Consistent with the earlier 2D work, the gas accretes near the midplane at a
super-Eddington rate while a strong, photon-driven outflow forms near the rotation axis. 
Outgoing mass flux appears only beyond $\sim 10\ r_g$ and can become comparable to the net accretion rate beyond $20$--$50\ r_g$; the inferred outflow depends on how it is defined~\citep{Yuan+2012}. For a non-spinning black hole the funnel luminosity saturates near $\Ledd$ at $\dot{M} \approx 100\,\Ledd/(0.057 c^2)$; for spin $0.9$ it rises with radius from $\sim\Ledd$ at $20\ r_g$ to $\sim 10\Ledd$
at $200\ r_g$, and the outflow rate is larger~\citep{Sadowski+2014}. The radiation driven outflow,
sometimes termed a ``radiative jet''~\citep{SadowskiNarayan2015}, has a
mass-weighted speed of only $0.2$--$0.3c$ due to radiation drag \citep{Jiang+2014a,TakahashiOhsuga2015}, quite distinct from the
ultra-relativistic Blandford--Znajek jet~\citep{DavisTchekhovskov2020}.

There are generally two types of disks as found by these simulations, depending on the choices of initial magnetic field topology and strength used in these simulations. One type is the standard and normal evolution (SANE), which has similar properties to the $\alpha$ disk model, except that the turbulence provides the self-consistent angular momentum transport. Thermal pressure typically dominates and the magnetic field never becomes strong enough to resist the 
inflow. Another type of solution is the magnetically arrested disk (MAD), 
where the accumulated flux becomes dynamically dominant. 
They are non-axisymmetric, show fast accretion-rate variability, and can launch relativistic jets magnetically~\citep{McKinney+2015,Yang+2023}.
Winds are driven by both radiation \emph{and} magnetic pressure, so MAD disks
produce more massive and faster winds than SANE disks. The wind is highly
anisotropic: ``ultra-fast outflows'' (UFOs) with $v > 0.1c$ surround the funnel,
with slower, denser winds farther out. Around fast-spinning holes,
a relativistic jet is launched by the Blandford--Znajek
mechanism~\citep{BlandfordZnajek1977}. The magnetized optically thick wind carries
radiation outward until it becomes optically thin, and the jet drills an
optically thin funnel that lets more photons escape. 

These strongly magnetized disks typically have larger inflow speeds and smaller surface densities than the SANE models with the same mass accretion rates. Consequently, MAD disks around fast-spinning holes can reach radiative efficiencies 
of $8\%-20\%$, which increases when the accretion rate rises from $\Medd$ to $\sim 10^3\Medd$ \citep{McKinney+2015, CurdNarayan2023}. This also implies that super-Eddington MAD disks around spinning black holes can have luminosity well above the Eddington limit. For non-spinning black holes, the radiative efficiency of super-Eddington disks is around $\sim 2-12\%$ with weak dependence on the accretion rate \citep{Sadowski+2016b,Kwan+2026}.

Simulations for super-Eddington accretion flows that solve for the full angular distribution of specific
intensities \citep{Jiang+2014,White+2023} have also been performed \citep{Jiang+2014a,Asahina+2020,Zhang+2025}, which are not subject to the artifacts of the M1 approach. Particularly in the funnel region, the M1 closure will merge photons emitted from different directions and can create artificial shock structures in the radiation field as photons are treated as a fluid in this approach \citep{Sadowski+2015,Sadowski+2016b}. 
There are two sets of simulations in this category depending on whether GR is solved or not. 
One set of simulations evolves a Newtonian flow, capturing GR effects around a Schwarzschild hole with a pseudo-Newtonian
potential~\citep{PaczynskiWiita1980}. Using this method, \cite{Jiang+2014a} simulated 
a disk accreting onto a $6.62\ \msun$ hole at  
$22\,\Medd$ (cylindrical coordinates, torus at $100\ r_g$). The inner
$80\ r_g$ radiates at $\sim 10\,\Ledd$ and the outflow carries $\approx 29\%$ of the
net accretion rate. A snapshot of the simulated accretion disk is shown in panel (a) of Figure \ref{fig:simulation_XRB}. A key result is that \textbf{magnetic buoyancy} enhances
cooling---low-density, magnetically buoyant regions rise and carry photons toward
the surface, reducing the effective optical depth so that photons need not
diffuse from the midplane. Similar behavior is found across accretion rates and
black hole masses in spherical-polar simulations~\citep{Jiang+2019,Huang+2023}, with
the funnel opening angle shrinking and the radiative efficiency falling to
$\sim 1\%$ by $\dot{M} \approx 150\,\Medd$~\citep{Jiang+2019}.
For supermassive black holes the ratio between radiation pressure and gas pressure is much larger ($\sim 10^3$--$10^5$
for $r<40\ r_g$ around a $5\times 10^8\ \msun$ hole, versus $10$--$10^3$ for
stellar-mass holes), making the flow far more compressible. This
produces clear spiral shock structures whose Reynolds stress becomes
comparable to or larger than the MRI Maxwell stress. No such spirals appear in
the stellar-mass case. Nevertheless, the radiative efficiency and outflow properties
at fixed Eddington ratio are largely insensitive to black hole mass.

\begin{marginnote}[]
\entry{Reynolds and Maxwell stress} {The Reynolds stress is the effective stress exerted on the mean accretion flow due to turbulent velocity fluctuation, while the Maxwell stress is the angular momentum flux carried by the magnetic fields. In normal MRI turbulence, Maxwell stress is typically larger than Reynolds stress.}
\end{marginnote}

GR schemes that solve the intensity equation directly have also been
developed~\citep{DavisGammie2020,White+2023}. A related approach uses the
intensities only to compute the Eddington tensor closing the moment
equations~\citep{Asahina+2020}. Compared to an axisymmetric GRMHD simulation
of a super-Eddington flow onto a non-magnetized neutron
star~\citep{Takahashi+2018}, this approach and M1 solutions agree near the optically
thick midplane but differ substantially near the surface and funnel, where M1 solutions
unphysically enhance the radial radiation flux near the axis~\citep{AsahinaOhsuga2022}.
A set of simulations \citep{Zhang+2025,Zhang+2026} to explore the properties of super-Eddington accretion disks based on the scheme developed by \cite{White+2023} have been performed with two spins $a = 0.3$ and $0.9375$, and two initial field topologies (a single loop carrying net vertical flux, and a double loop carrying none). These simulations all use simple opacities scaled to stellar-mass black holes. They are all SANE models even though magnetic pressure can dominate over gas and radiation pressure in certain regions of the disk. The runs with net vertical magnetic flux
and rapid spin still launch powerful relativistic jets, which are shown in panel (b) of Figure \ref{fig:simulation_XRB}. The radiative efficiency (measured at $20\ r_g$) falls monotonically from
$5.4\%$ near and below Eddington to $0.45\%$ at $\dot{M}/\Medd = 149$, which is also quite consistent with the results from the runs using the Pseudo-Newtonian potential.
Radiation generated in the inner
disk drives substantial outflow that shapes a narrow, conical funnel-shaped
photosphere. These simulations show three distinct regions of the disk: the jet with high Lorentz factors, the wind with typical speed $0.2c-0.4c$, and an optically thick disk at the disk midplane. The disk becomes thicker and squeezes the jet and wind regions when the accretion rate is increased. 

All the simulations discussed above start with a rotating torus for simplicity, which is also a classical approach for global MHD simulations of accretion disks \citep{Hawley2000}. The disks are also aligned with the direction of spin of the black hole. \cite{Fragile+2025} adopt a different initial condition for their 3D general relativistic radiation MHD simulations based on the M1 closure. An equilibrium disk structure is constructed based on the GR generalized thin disks \citep{NovikovThorne1973} up to the radius $30\ r_g$. 
Unlike the cases starting from a rotating torus, 
these simulations start with an initial condition for a constant mass accretion rate constructed to be $1-10\Medd$ with a fixed $\alpha=0.02$. The net accretion rates onto the black hole all settle to $\Medd$ with large inflow and outflow rates almost canceling out. This result is quite different from previous simulations that start with a rotating torus and it clearly demonstrates the importance of initial conditions for these simulations, although it is unclear which initial condition is closer to reality.

If the initial Keplerian disk is tilted with respect to the black hole spin, the mass accretion rate is enhanced and can exceed the Eddington value \citep{Fragile+2026}. This is caused by 
two standing (opposing) shocks in the inner region (a few $r_g$), which are driven by the latitude-dependent radial epicyclic response of tilted orbits. The shocks extract angular momentum and shorten the residence time in the inner disk. This mode of accretion enhances the mass accretion rate without significant dissipation, which reduces the overall radiative efficiency.

Due to the different numerical schemes, initial conditions, and magnetic field topologies adopted, numerical simulations for the inner regions of super-Eddington accretion disks performed by different groups do not agree with each other exactly. 
However, they do share some common properties.  
\begin{itemize}
	\item \textbf{Radiative Efficiency}.
The radiative efficiencies reported by these simulations are summarized in Figure \ref{fig:efficiency}. We have converted different definitions of $\Medd$ adopted in the original published papers to the uniform definition $\Ledd/\left(0.1c^2\right)$.  
Most of the MAD runs show radiative efficiencies orders of magnitude larger than the non-MAD runs due to lower surface densities as well as larger funnel regions. For the non-MAD runs, the radiative efficiencies generally decline with increasing mass accretion rates as in the standard semi-analytical model for super-Eddington accretion, although the exact value can vary. 
As we will discuss later (\S \ref{sec:theo_summary}), the observed properties of super-Eddington sources can provide constraints to distinguish different accretion disk models. 

 \item \textbf{Mechanical power of the outflow}.
Outflow is also an unavoidable product of super-Eddington accretion for all these simulations. Calculations that adopt general relativity can produce relativistic jets along the rotation axis with non-zero spin in addition to the wind with a larger opening angle. The mechanical luminosities carried by the jet and outflow as found by different simulations are summarized in Figure \ref{fig:kin_efficiency}. 
The jet power is set by both spin and accretion rate. The two MAD models from \cite{CurdNarayan2019}, labelled m00 ($a=0$) and m09 ($a=0.9$),
differ by a factor of $100$ even though the non-spinning one is accreting
faster, and the \cite{CurdNarayan2023} $a=0.9$ models lie an order of
magnitude above their $a=0$ counterparts. The wind power instead grows steadily with accretion rate and
depends far more weakly on spin, though at fixed $\dot{m}_{10}$ it still
spans two orders of magnitude between the MAD and non-MAD runs. Note that
the radius at which the outflow is measured is not uniform, ranging from
a few hundred to a few thousand $r_g$, and radiatively driven outflows
keep gaining kinetic energy out to and beyond the wind photosphere.

	\item \textbf{Geometric beaming}.
	Geometric beaming is a universal outcome of super-Eddington accretion
	simulations that form a radiation-pressure-inflated envelope with an
	evacuated polar funnel. The luminosity weighted beaming factor $b$ is usually defined as the ratio between the intrinsic luminosity and the equivalent isotropic luminosity given the flux at a particular viewing angle. This is closely related to the opening angle of the funnel, which narrows with increasing accretion rates. It can vary from $55^{\circ}-37^{\circ}$ when $\dot{M}/\Medd=1.3-4.2$ \citep{Mills+2024} to a beam of $\approx 20^{\circ}$ when $\dot{M}/\Medd=10$ \citep{Narayan+2017}. 
	The implied beaming factor is moderate, e.g., $\sim 0.1-0.5$ across different simulations when $\dot{m}/\Medd \lesssim 10$ \citep{Jiang+2014a,Jiang+2019}. However, when $\dot{m}/\Medd > 100$, MAD simulations predict a beaming factor of $10^{-3}$ due to the relativistic jet \citep{Narayan+2017,CurdNarayan2019}, which is too strong to be consistent with the nebulae observations of ULXs as we will discuss later.
	
	\item \textbf{Caveats}. All these simulations start with an idealized initial condition and only run for a limited amount of time. The results may depend on these choices. Inputs from large scale properties of the host galaxy are needed to close the gap. 
	
\end{itemize}

\subsubsection{Spectra from the Simulated Super-Eddington Accretion Disks}
\label{sec:sim_spec}

When the accretion rate increases from the sub-Eddington to super-Eddington regime, the changes in disk structure determine the evolution of the intrinsic spectra. Take the supermassive black hole case as an example. The commonly observed X-rays for Type 1 AGNs are from the inner corona region, which are produced when a significant fraction of accretion power is dissipated in the optically thin region \citep{Jiang+2014b,Jiang+2019b}. 
In the optically thick part of the disk, once the gas and radiation are thermalized, the temperature will be too low to produce any hard X-ray photons via inverse Compton.  
This ultimately depends on the surface density of the disk, which decreases towards the black hole and increases with mass accretion rate in global simulations of accretion disks with self-consistent MRI turbulence. For typical super-Eddington disks, the surface density is large enough that the disk is optically thick as discussed in section \ref{sec:model}. Therefore, the corona and the associated X-ray emission get weaker with increasing accretion rate. This is also the physical reason for the $\alpha_{\rm OX}$ relation observed for AGNs, where the ratio between optical flux and X-ray flux increases with increasing luminosity. Notice that the trend is less clear for MAD disks, which can still produce strong X-rays even with super-Eddington accretion rates and are inconsistent with the observed spectrum properties.

The only place that is still optically thin that allows hard X-ray photons to be produced by thermal and bulk Compton processes is the funnel region in super-Eddington disks. The resulting spectrum is typically non-thermal and can show up as a power law extending to the hard X-ray band \citep{Kitaki+2017,Narayan+2017,Mills+2024}. X-ray binaries and AGNs can share many similarities for this part of the spectrum as photons are not thermalized and the flow velocities for bulk Compton are not sensitive to the black hole mass. Because the funnel region has a small opening angle, which also narrows as accretion rate increases, the possibility to observe the hard X-rays will be viewing-angle dependent. Face-on observers that can see through part of the funnel are able to detect hard X-rays while observers with larger inclination angles will not. 

The resulting spectra for super-Eddington disks are likely a superposition of several components: a non-thermal spectrum from the funnel region, emission from the photosphere of the puffed-up disk, or the optically thick wind, as well as low energy emission from the outer part of the disk. Thermal emission from the photosphere is typically in the soft X-ray or EUV bands for X-ray binaries, while for AGNs, it will be in UV/Optical bands. This is how the spectral model for narrow-line Seyfert 1 galaxies is constructed by \cite{KubotaDone2018}, where the slim disk model is used to describe the puffed-up disk while a standard thin disk is used to model the outer part of the disk. 

The only missing piece is the soft X-ray excess (see definition in \S \ref{sec:obs_summary}). The soft X-ray excess may be a high-energy extrapolation of the far-ultraviolet (FUV) spectrum beyond $1000$~\AA\ \citep{Laor+1997,Zheng+1997,Czerny+2003}, as the continuum FUV and soft X-ray spectra of radio-quiet quasars are consistent in both slope and amplitude. 
Two phenomenological models have been proposed to describe the soft X-ray excess. One is relativistically blurred ionized reflection, in which the ``excess'' is the
smeared blend of soft emission lines from the disc surface \citep{Crummy+2006}. The other assumes a separate \emph{warm corona} component, which is an optically thick ($\tau\sim 40-50$), $kT_{\rm e}\sim0.1$--$0.2$~keV
Comptonizing layer covering the inner disc 
\citep{Jin+2012,Petrucci+2020}.
The parameters are tuned to fit the data without any physical connection to the underlying properties of the disk. 

An alternative physical mechanism to produce the soft X-ray excess is proposed by \cite{Jiang+2025} based on three-dimensional radiation MHD simulations of
accretion disks around a $10^{8}\,\msun$ black hole. These simulations focus on the radial range $\sim 50-200\ r_g$. Radiation spectra are calculated self-consistently based on multi-group calculations, which peak near $10$~eV as expected for this mass. In addition, the models at and above roughly the Eddington rate
develop a power-law continuum between $\sim10$~eV and $1$~keV with
the slope $-1$ to $-2$ comparable to the radio-quiet quasar slopes quoted above.  
Because the super-Eddington disks are thick, the inflow speed $v_r$ of the gas is larger than the electron thermal speed. The thermal photons produced by the main body of the disk are boosted by \emph{compressible bulk Comptonization} in the converging accretion flow over the optical depth range of $c/v_r$, which explains why the soft X-ray excess is the continuum extension of the FUV spectrum. 
The photon energies come from bulk motion rather than from the local temperature. Therefore the properties of the soft X-ray excess are not sensitive to the thermal properties of the gas, although this still needs to be demonstrated for different black hole masses. 

None of the simulations discussed in section \ref{sec:sim_inner} covers the outer part of the disk to address the question of whether the low frequency part of the spectrum can be described by a standard thin disk or not. \cite{Narayan+2017} performed a series of 2D and 3D simulations of super-Eddington accretion disks onto a 10 $\msun$ black hole and calculated the low frequency spectra from these simulations via the post-processing approach based on a ray-tracing scheme. However, this is achieved by extrapolating the disk profiles to a much larger radius than the radial range where a steady state is achieved for these simulations. Future simulations are needed to explore the global structures of super-Eddington accretion disks over a wide radial range self-consistently.

\subsubsection{Variability of super-Eddington Accretion Disks from the Simulations}
\label{sec:sim_variability}
Variability at different wavelengths is controlled by three different characteristic timescales in the disk: the local dynamical timescale $t_{\rm dyn}$, the thermal timescale $t_{\rm th}$ and the inflow timescale $t_{\rm inflow}$. They are defined as 
\begin{align}
	t_{\rm dyn} =&\ 0.18 \ {\rm d}\
\left(\frac{M}{10^{8}\msun}\right)
\left(\frac{R}{10\ r_g}\right)^{3/2}, \\
t_{\rm th} =& \left(\frac{1}{\alpha}\right) t_{\rm dyn}
=
1.80\ {\rm d} \left(\frac{\alpha}{0.1}\right)^{-1}
\left(\frac{M}{10^{8}\msun}\right)
\left(\frac{R}{10\ r_g}\right)^{3/2}, \\
t_{\rm inflow} =& \left(\frac{H}{R}\right)^{-2} t_{\rm th}
=180\ {\rm d} \
\left(\frac{\alpha}{0.1}\right)^{-1}
\left(\frac{H/R}{0.1}\right)^{-2}
\left(\frac{M}{10^{8}\msun}\right)
\left(\frac{R}{10\ r_g}\right)^{3/2}.
\label{eq:inflow_time}
\end{align}

If the radiation field is thermalized and the effective temperature varies with radius as 
$T_{\rm eff}\propto R^{-\beta}$, the radius that dominates the emission at wavelength $\lambda$ will scale as $R(\lambda)\propto \lambda^{1/\beta}$ due to Wien's displacement law. Therefore, the characteristic timescales naturally increase with $\lambda$. Another factor that affects the variability timescale and amplitude is the optical depth. X-rays in AGNs as well as the hard X-rays in X-ray binaries are produced from the optically thin inner regions of the accretion flow. Therefore the X-ray can vary at the local dynamical timescale with large amplitude, which can be sub-day for AGNs and $\sim 10^{-3}$ s for X-ray binaries. UV, optical, and infrared photons are produced at much larger radii and the disk there is optically thick. The variability amplitude is generally reduced by the photon random walk through the optically thick disk. Their variability timescales will be months or years for AGNs.

These general trends of variability in AGNs still hold for super-Eddington sources. In addition, there are two other effects in super-Eddington disks that affect the variability. 1) The puffed up inner region of super-Eddington accretion disks will obscure the inner X-rays, which will reduce the UV/Optical variability due to X-ray irradiation. 2) Outflows produced in super-Eddington accretion disks will also push the photosphere to much larger radii and thus reduce the variability amplitude and increase the characteristic variability timescale. This picture is generally consistent with the observed properties of UV/Optical variability as a function of Eddington ratio \citep{Zuo+2012,Simm+2016,Yu+2022}. 
This is also basically what is found by \cite{Zhang+2025} that super-Eddington disks generally vary less than the sub-Eddington disks. The fractional variability amplitude of the radiation field emitted from the optically thick photosphere is also generally much smaller than the fractional variability of the mass accretion rate itself due to the smearing effect of large optical depth \citep{Jiang+2019}.

There are exceptions to this general trend of variability. For typical conditions expected for AGN disks in the UV/Optical emission region, the dominant opacity that determines the coupling between gas and radiation is the bound-bound transitions from metals, particularly the iron group elements. This opacity is most important for temperatures around $\sim 2\times 10^5$ K and densities $\sim 10^{-8}-10^{-12}$ g/cm$^3$, which are very similar to the conditions expected in the envelopes of massive stars \citep{Jiang+2016}. This opacity can enhance the Rosseland mean opacity by a factor of few compared to the electron scattering value. When the accretion rate is above the Eddington value, as defined by the electron scattering opacity, the radiation force caused by the enhanced opacity is already much larger than gravity. This will produce strong convection in the main body of the disk and cause the disk to oscillate. The local radiation flux can vary by a factor of a few on the local thermal time scale \citep{JiangBlaes2020}. The physical mechanism for the oscillation is similar to the kappa mechanism that is responsible for stellar oscillations, which is due to the temperature dependence of the opacity. Large opacity will cause the local temperature to increase. Once the gas becomes hotter than $\sim 2\times 10^5$ K, the opacity will drop and cooling is enhanced. The temperature will then drop and the opacity peak will show up again. The disk thickness will also oscillate when convection is operating, which will produce variable obscuration of the central region for edge-on observers. 

For MAD disks, magnetic fields will build up and erupt periodically roughly every $\sim 2000 G\mbh/c^3$ \citep{CurdNarayan2023}, which corresponds to $11.6$ days for a $10^8\msun$ black hole. This will cause quasi-periodic modulation of the radiation field with luminosity changes by more than $25\%$. This quasi-periodic magnetic flux eruption typically causes the MAD disks to show much larger variability amplitude than SANE disks, which has been studied in great detail for the radiatively inefficient low accretion rate regime \citep{ChatterjeeNarayan2022}.

One important caveat to all of these simulations is the limited dynamical range and duration of the simulations, which makes it impossible to test the variability properties over a much longer timescale. \cite{Secunda+2026} 
extrapolates the variability properties of the simulated AGN disks around $\sim 100\ r_g$ \citep{Jiang+2025} to $\sim 10^5\ r_g$ as needed for little red dots. It is generally consistent with the current data and works better than naive expectation based on the standard thin disks. Future simulations will be needed to test it directly.

\subsection{Uncertainties at the Mesoscale}
\label{sec:mesoscale}
All the simulations discussed in section \ref{sec:sim_inner} need to start with the gas and magnetic field relatively close to the black hole ($\lesssim 100 \ r_g$) either in the form of rotating tori or a prescribed disk-like structure. The magnetic field topology, strength and the angular momentum budget of the initial gas do affect the properties of the disk significantly. In principle, for supermassive black holes, simulations of isolated galaxies can provide enough information on the properties of gas and magnetic fields that will feed the black hole. For X-ray binaries, properties of the companion, either in the form of winds or Roche lobe overflow, determine the supply to the disk. However, very few simulations have shown how accretion happens at the mesoscale, $\sim 100\ r_g - 10^6\ r_g$. For AGNs, this corresponds to $r<0.1$ pc, roughly below the resolution of typical galaxy scale simulations. 

If the accretion disk is in steady state as commonly assumed, the disk at the mesoscale is expected to be geometrically thin even at super-Eddington accretion rates. For supermassive black holes, self-gravity starts to become important at this mesoscale due to low temperature and high surface density. \cite{Sassano+2023} studies super-Eddington accretion onto a $10^3\msun$ black hole seed at redshift $15$. The Toomre $Q$ parameter quickly drops below unity and a self-gravitational instability kicks in, which makes the disk highly fragmented and prone to forming stars. This conclusion is robust to the detailed sub-grid model adopted by Sassano et al.\ for radiative feedback from the super-Eddington accretion disks. The net result is that super-Eddington accretion only operates for a short period of time ($\sim 0.1-0.3$ Myr) before the accretion rate drops below the Eddington value. 

The self-gravity barrier can be avoided if the disk is strongly magnetic-pressure dominated. \cite{Hopkins+2024a} finds that the whole super-Eddington disk from the inner region to the mesoscale is dominated by magnetic pressure, because magnetic fields are advected with the gas via the flux frozen condition. The surface density is significantly reduced in such disks as discussed in section \ref{sec:model}. Therefore, the self-gravitational instability does not kick in. However, the inner regions of such strongly magnetic-pressure--dominated super-Eddington disks also have much smaller surface densities compared with the simulations discussed in section \ref{sec:sim_inner}. The X-ray emission will be stronger and the variability amplitude will be larger for such disks, which may not be consistent with the observational properties of super-Eddington sources as discussed in section \ref{sec:obs_summary}. 

The commonly used steady-state assumption may not be true at the mesoscale. According to equation \ref{eq:inflow_time}, the inflow time will be $\sim 16$ Myr at $10^6\ r_g$ for a $10^8\msun$ black hole. Galaxy-scale simulations including feedback also find that super-Eddington accretion can only last for a short period of time with a duty cycle on the order of a Myr. Therefore, it may be common for the accretion rate at the mesoscale to be different from the accretion rate at the inner region. If the accretion rate is larger at the mesoscale, the disk there will become significantly thicker while the inner region remains thin. Such thickening of the outer disk may produce more outflows on large scales, such as those observed in ERQs and HotDogs.

%

The properties of accretion disks at the mesoscale could be probed with GRAVITY+ in the near future \citep{Abuter+2022}. 
The near-infrared interferometric astrometry will have enough spatial resolution to resolve 
the sub-parsec broad-line region in hundreds of AGN, mapping their size, geometry, rotation and outflow kinematics directly. Spectropolarimetry will also be a powerful tool that can constrain the dynamics of the flow (such as disk thickness and outflows) at these scales \citep{Alexandroff+2018,Zakamska+2023}. Improved measurements of gas at the mesoscale will also provide valuable information about the proper initial and boundary conditions to use for numerical simulations of super-Eddington accretion disks, which is one of the biggest uncertainties for simulations as discussed in \S \ref{sec:sim_inner}.

\subsection{Super-Eddington Accretion in galaxy simulations}

Super-Eddington accretion is one promising way to grow the black hole mass efficiently in the early universe. Although the typical resolution of current galaxy scale simulations is not sufficient to resolve the inner regions of the accretion disk around the black hole, the accretion disks and their host galaxies are strongly connected. On the one hand, the radiative and mechanical feedback from the disk can have a big impact on the environment of the black hole at a galactic scale. On the other hand, gas and magnetic fields in the galaxy determine the initial and boundary conditions for the accretion disk, which can have a big impact on the disk itself. Current galaxy scale simulations are trying to answer the following questions for super-Eddington accretion. 
\begin{itemize}
	\item Is there a large enough gas reservoir to fuel super-Eddington accretion? This means the black hole needs to accrete with a rate larger than the Eddington mass accretion rate $\approx 2.2~(\mbh/10^8\msun)\ \msun/{\rm yr}$.
	\item How does the gas transport angular momentum efficiently so that it can reach the black hole? 
	\item Will the radiative and mechanical feedback from the black hole stop the accretion on large scales?
	\item What is the duty cycle of super-Eddington accretion?
\end{itemize}

\begin{figure}[htbp]
	\centering
    \includegraphics[width=\linewidth]{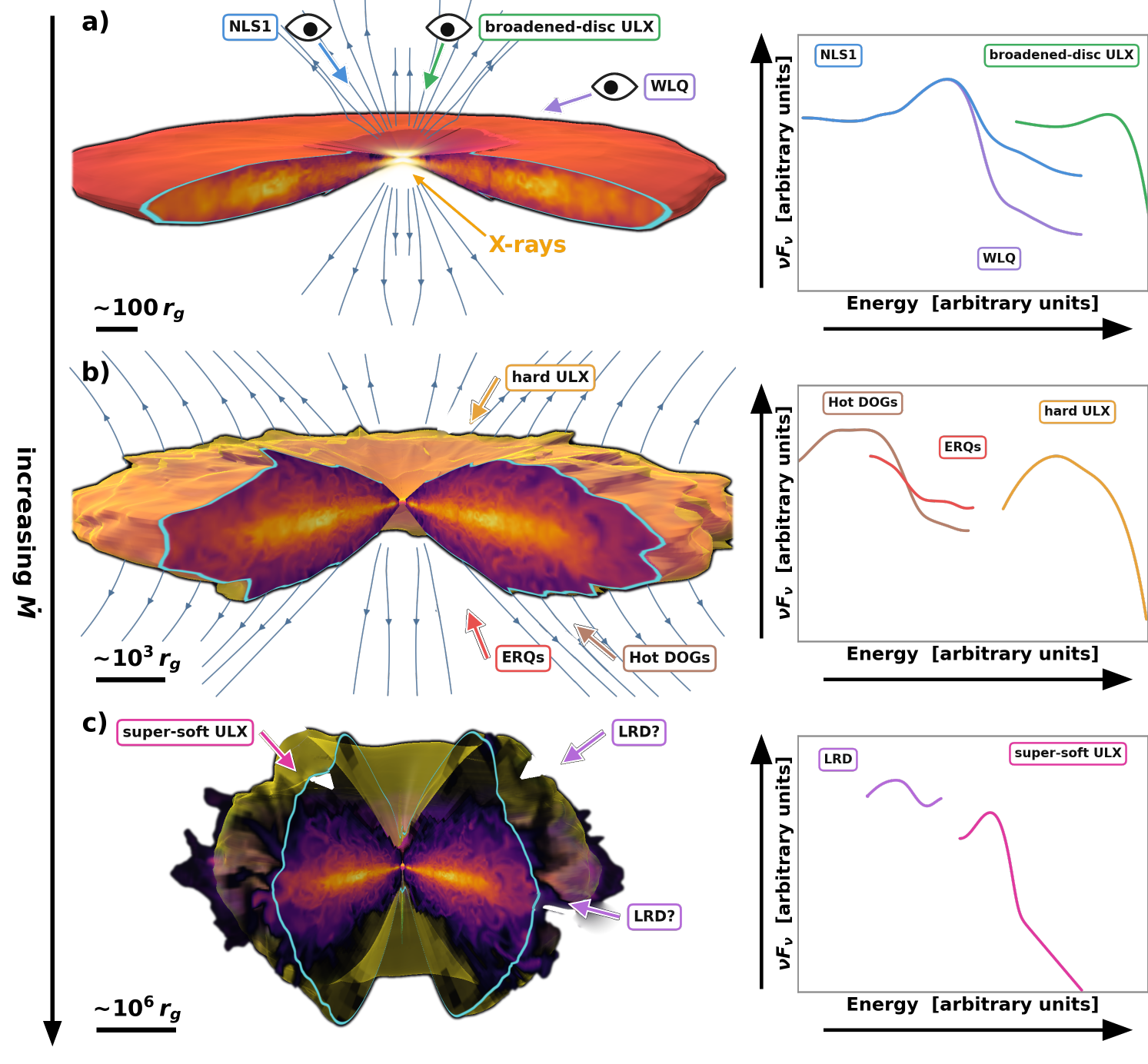}
		\caption{A cartoon of the changes of super-Eddington accretion disks as a function of increasing mass accretion rate (a) to (c), indicating how the observed spectra (right) depend on both mass accretion rate and viewing angle. The scale bar in each row gives its
	approximate size, from $\sim100\,r_g$ in (a) to $\sim10^{6}\,r_g$ in (c). 
	\emph{Left:} a meridional cut through the flow, showing the turbulent
	main body of the disk with the photosphere indicated by the cyan lines. The outflows launched from the disk are indicated by the blue streamlines. Different astrophysical sources are associated with each stage at the viewing angles from which they may be seen. 
	\emph{Right:} schematic $\nu F_\nu$ spectra (arbitrary units) for the three stages with corresponding samples. }
	\label{fig:cartoon}
\end{figure}

Feedback is crucial to regulate super-Eddington accretion on galactic scales. However, in detail the impact depends on how it is implemented in numerical simulations, which is one of the biggest uncertainties in connecting galaxy scales with the physics at the scale of the accretion disk. Radiative feedback can have two different forms. The first is energy driven, which is typically implemented as an energy source term added to a sphere of choice isotropically as $\dot{E}=\epsilon_f \eta_r \dot{M}c^2$. Here $\eta_r$ is the radiative efficiency of the accretion disk for a given mass accretion rate $\dot{M}$, while $\epsilon_f\approx 0.05-0.15$ is an empirical factor to convert the radiative luminosity to the local energy source term of the gas \citep{Springel+2005,BoothSchaye2009}. If the gas can cool efficiently, radiative feedback implemented in this way may not have any significant impact. In practice, stochastic heating with large temperature fluctuations is needed to avoid the issue of fast cooling \citep{DallaVecchiaSchaye2012}. The radiative efficiency $\eta_r$ roughly follows the results summarized in Figure \ref{fig:efficiency} for the non-MAD case. It decreases with increasing mass accretion rate and thus the radiative feedback becomes less important in these simulations \citep{Lupi+2016,Koudmani+2024}. 

The second form of radiative feedback is momentum-driven, where the radiation force is added to the momentum equation of the gas as $\tau\Ledd/c$, where $\tau\sim 10$ is taken to be the infrared trapping factor. This model of radiative feedback is not sensitive to the issue of cooling but the momentum of photons $\Ledd/c$ is typically small compared with the momentum budget in the system. Thus, this mode is only considered for the radiation force on dusty gas \citep{IshibashiFabian2015}.

In cosmological simulations, mechanical feedback by the wind is implemented via a random velocity kick to the gas with the kinetic energy flux normalized to be a fixed fraction of $\dot{M}c^2$ \citep{Weinberger+2017,Husko+2024}. More common is mechanical feedback via a jet, which determines the energy flux as $\dot{E}_{\rm jet}=\eta_{\rm jet}\dot{M}c^2$ with $\eta_{\rm jet}=1.3 a_{\star}^2\phi_{\rm MAD}^2$. This is based on analytical jet models  \citep{BlandfordZnajek1977,DavisTchekhovskov2020} and is also consistent with numerical results for the jet power as shown in the top panel of Figure \ref{fig:kin_efficiency}. The jet can deposit energy \citep{Sijacki+2007}, momentum \citep{Husko+2022} or both \citep{Dubois+2012,Massonneau+2023} into the gas at galaxy scales. Jet feedback is typically the dominant process for super-Eddington accretion because the jet energy efficiency, unlike the radiative efficiency, does not depend on the mass accretion rate. However, it has a strong dependence on the black hole spin as well as the poloidal magnetic flux at the horizon scale, both of which are sub-grid model parameters. \cite{Lupi+2024} is able to sustain super-Eddington phases for a much longer period of time for a fixed spin value ($0.7$) to avoid the strongest feedback from maximumally spinning black holes. 

The current galaxy scale simulations can be improved in several ways to tell a more consistent story of the growth of black hole seeds via super-Eddington accretion. 
Figure \ref{fig:kin_efficiency} suggests that mechanical feedback from a wind is both more common and more robust. The energy flux carried by the wind is as large as that carried by the jet but, as discussed in \S \ref{sec:radio}, jets are not commonly observed for super-Eddington sources. More importantly, the mechanical output from the wind is not sensitive to the very uncertain evolution of black hole spin and poloidal magnetic flux. The wind also has a larger opening angle compared with the jet, which will allow it to couple with the surrounding gas more efficiently. The detailed processes that couple the radiation, wind and jet produced by the accretion disks with the galaxy are still poorly understood. 


\section{Connecting the Simulations and Observations of super-Eddington Accretion}
\label{sec:theo_summary}
Semi-analytical models and numerical simulations have successfully explained many observed properties of 
super-Eddington sources, including unique features in their spectra, variability and outflows. However, 
there are still big gaps to connect the two aspects of super-Eddington accretion. In this section, we are going to summarize 
the successes and point out the challenges for the existing simulations to explain the observations. Before that, we are going to first summarize the differences between the thermal pressure and magnetic pressure dominated super-Eddington accretion disks found by different simulations to clarify the simulation results we focus on. 

There are two types of accretion disk models found by simulations in the super-Eddington regime. For the first type, the disk is dominated by radiation pressure and the disk properties share many similarities with the classical semi-analytical models. Gas and radiation are typically thermalized as the flow is optically thick. The outflow is also optically thick, as in the model of \cite{Poutanen+2007}. These are the models we will focus on because they produce many properties that are consistent with the data, as we will explain below. The other type of disk is dominated by magnetic pressure, either in the form of MAD disks or the strongly magnetically elevated disks. Compared with the thermal pressure supported disks with the same mass accretion rates, these disks have smaller surface densities and larger inflow speeds. They more easily produce powerful jets and they emit more X-ray photons. They are not necessarily X-ray weak even with super-Eddington accretion rates. The spectrum is also hard over a large range of viewing angles ($\sim 30^{\circ}$). On the other hand, these magnetically pressure-dominated disks can avoid the issue of fragmentation due to self-gravity as discussed in \S \ref{sec:mesoscale}, which is a potential issue for thermal-pressure--dominated disks. 

%

\subsection{Overview of model-data comparison}

We will first attempt to synthesize the common properties across different super-Eddington objects from \S \ref{sec:obs_summary} with the theoretical expectations laid out in \S \ref{sec:model}. We suggest that the observed phenomenology that is associated with super-Eddington accretion is best thought of in a two-dimensional way, where properties evolve both as a function of Eddington ratio and viewing angle. We summarize the progression in Figure \ref{fig:cartoon}. 

\begin{itemize}
\setlength{\parindent}{15pt}
\setlength{\itemsep}{6pt plus 2pt minus 1pt}

    \item \emph{Similarities and Differences between super-Eddington accretion onto stellar mass and super-massive black holes.} 
Typical temperatures in the super-Eddington disks around the stellar mass and super-massive black holes are quite different as the temperature scales with black hole mass as $\mbh^{-1/4}$. However, their general spectral shapes share many similarities as shown in the two panels of Figure \ref{fig:sed_cartoon}. The hard X-rays (above $\sim 5$ keV) in X-ray binaries are thought to be produced by the central corona region. When the accretion rate increases to the broadened disk and ultraluminous stages, the spectral slopes get steeper. When the X-ray region is fitted with a comptonization model, a higher optical depth is needed, which implies that the density in the central corona region is higher, as expected for super-Eddington accretion. The corona region may not exist in the supersoft state. This is consistent with the X-ray weakness for AGNs as shown in Figure \ref{fig:sed_cartoon}, as well as a softer X-ray spectrum for the super-Eddington AGN sources discussed in \S \ref{sec:obs_summary}.

In ULXs, most of the photons produced by the main body of the disk are still in the soft X-ray band and the spectral shapes are not exactly the same as the UV/Optical spectra of super-Eddington AGNs. Part of the reason is due to the different physical processes that are operating in the two systems. In X-ray binaries, thermal Compton is more important due to higher gas temperature, while in AGNs, only bulk Compton can potentially play an important role for the soft X-ray excess. The local gas opacity is also larger for AGNs because of numerous bound-bound and bound-free transitions in metals that are not relevant for X-ray binaries. This will make AGN disks thicker and produce cooler outflows. The UV/Optical lines that are typically used to trace outflows in AGNs do not exist for X-ray binaries. Dust and molecular gas that are important for the properties of ERQs and Hot DOGs also do not exist for X-ray binaries. But the central X-ray region for the two, when it exists, shares many common properties as discussed in \S \ref{sec:variability}. 

It is also interesting to compare LRDs with the supersoft state of ULXs. They both show blackbody like spectra and can be interpreted as emission from photospheres at larger radii from the central black holes. Corona emission is extremely weak for both of them. Other spectral features in LRDs, such as the Balmer break and the absorption and emission lines, do not exist in the supersoft ULXs due to the vastly different gas conditions. It is also interesting to note that the supersoft ULXs can change to a typical broadband ULX state in the timescales of hours to days \citep{UrquhartSoria2016,DAi+2021}. These timescales are consistent with the inflow timescales at $\sim 10^3-10^4\ r_g$. It is generally interpreted that when the accretion rate drops and the photosphere size is reduced below $10^3\ r_g$, the hard X-rays start to show up. It is also believed that LRDs have a duty cycle of 1-10 Myr \citep{Arita+2025,Sun+2026}, although we cannot observe the whole cycle directly. 
This is the reason we think they may belong to the same evolution stage as shown in Figure \ref{fig:cartoon}.

    \item \emph{Objects in three different broad categories.}
If organized by Eddington ratio and orientation, we can start with systems that are likely radiating around the Eddington limit. In this group are sources like the narrow-line Seyfert 1 galaxies and some of the weak-line QSOs. In these sources, we see the UV emitting part of the disk, and typically we see X-ray emission. Often the soft X-ray is highly variable -- these are likely cases where we are looking down the funnel region. There is also an X-ray weak tail to these sources -- likely these are the highest Eddington sources, and we are not seeing down the funnel, but rather observing the wind. The broad permitted \ion{Fe}{ii} emission that typifies EV1 emerges from outside the wind region, and is presumably enhanced because of X-ray shielding by the wind. 

As we move up in Eddington ratio to the middle panel of the figure ($\dot{M}/\dot{M}_{\rm Edd} \sim 2-10$), the photosphere of the inner wind-dominated flow moves outwards in radius, and the funnel region gets narrower. Most of the time now, observers will see sources like HotDOGs or ERQs. The very high velocity [\ion{O}{iii}] likely emerges from the outer parts of the wind. However, we cannot see the inner accretion flow through the wind, so UV/optical emission and variability are both suppressed. The small fraction of the time we are looking down the funnel, we observe very UV-luminous but X-ray weak sources with weak UV lines. 

Finally, in the bottom panel of Figure \ref{fig:cartoon}, we show the most extreme sources, at the highest Eddington ratios. At this point, the wind photosphere has grown comparable in size to the optical-emitting portion of the disk, and so all we see is the wind photosphere. Speculatively, we posit that both Super-soft X-ray sources and LRDs might be in this category as we have explained.
%

\end{itemize}

Overall, we expect X-ray weakness, low optical/UV variability, and vigorous outflow as generic properties of super-Eddington sources over this full range. In the subsections below, we examine detailed predictions and compare with observations.

\subsection{Nebulae around ULXs constrain the energy output from super-Eddington accretion}

Several nearby ULXs sit at the center of extended optical nebulae, which are typically $30-400$~pc
across, discovered in narrow-band imaging and long-slit spectroscopy
\citep{PakullMirioni2002,Kaaret+2004}. They can be ionized either by shocks or X-ray photoionization. The low-ionization spectra of [\ion{S}{II}] and [\ion{O}{I}] and the expansion velocity of the nebulae are typically used to infer the time averaged mechanical power. For IC 342 X-1, the average mechanical power is constrained to be $\sim 3.4\times 10^{39}$ erg/s \citep{Cseh+2012}, which is about $20\%$ of the X-ray luminosity. This value can be consistent with Figure \ref{fig:kin_efficiency} for the non-MAD models if the accretion rate is a few times the Eddington value. For nebulae dominated by recombination caused by the source's own EUV and soft X-ray photons, the typical diagnostic is the \ion{He}{II} $\lambda4686$ recombination line, as well as the mid-infrared [\ion{O}{IV}] $25.89$ $\mu m$, which requires photons above $54.4$ eV. Since the emission from the nebula is isotropic and sees the bolometric luminosity from the accreting system, by comparing the ionization luminosity required to power the nebulae with the observed luminosity from the source, we can constrain the beaming factor. Photoionization models for Holmberg II X-1 and IC 342 X-1 suggest that the inferred photoionizing luminosity is $\approx 3-6\times 10^{39}$ erg/s, which is only a factor of few smaller than the observed isotropic-equivalent X-ray luminosity \citep{PakullMirioni2002,Cseh+2012}. The small factor suggests that X-ray emission is only mildly beamed, which is inconsistent with the strongly beamed simulation results discussed in section \ref{sec:sim_inner}. One exception is NGC 5408 X-1, where simultaneous modeling of \ion{He}{II} and \ion{He}{III} regions requires the ionization luminosity to be smaller than the observed isotropic equivalent EUV and X-ray luminosity by a factor of 24 \citep{Guo+2026}. 

There are several caveats to keep in mind when comparing the nebulae. The EUV and soft X-ray photons needed for the photoionization may not be the same X-ray luminosity from the funnel region as typically determined from simulations. There can be significant contribution from the photosphere of the wind, which is not beamed at all. 
Detailed ionization modeling to convert the nebulae emission to the ionization luminosity is also subject to the uncertainty of the radiation spectrum from the source.

\subsection{Variability and Reverberation Mapping}
The impact of super-Eddington disks on reverberation mapping signals has been studied in numerical simulations. 
\cite{Thomsen+2022} computed X-ray reverberation signatures using ray-tracing 
based on the flow structure as determined by a super-Eddington radiation GRMHD simulation. 
Coronal photons are mostly reflected by the fast and optically thick winds rather than by a thin disk. 
The resulting response function is narrow and peaked at a light crossing time of $\sim 10\ r_g$. The 
iron $K\alpha$ line is also symmetric rather than skewed as typically found for reflection from a thin disk. 

\cite{Secunda+2026b} studied the impact of X-ray irradiation on the variability of UV/Optical photons based on multi-frequency radiation MHD simulations. When the disk thickness is small, X-rays emitted from the corona at the center can propagate through the disk on a light crossing time and they are the main driver of variability in UV-optical light curves. However, when the disk gets thicker, most X-rays get obscured and re-emitted in the longer wavelength bands. The correlations between X-rays and UV-optical lightcurves become weaker and the lags are $3-5\times$ longer. These results are in good agreement with numerous recent observations \citep{Guo+2022,Steyn+2026}. 

These simulations reveal two main effects in super-Eddington accretion disks that will change the variability properties compared with their sub-Eddington counterparts as we discussed in \S \ref{sec:sim_variability}. The first effect is the X-ray weakness and the lack of X-ray irradiation. As the disk gets thicker and the corona gets weaker, X-ray irradiation becomes less important and the UV/Optical variability will be dominated by the local turbulence \citep{Secunda+2025,Secunda+2026b}. The second effect is the optically thick outflow, particularly when the accretion rate is significantly above the Eddington limit. This will push the photosphere away from the disk, reduce the variability amplitude, and increase the variability timescale (see the cartoon shown in Figure \ref{fig:cartoon}). For observers looking through the funnel region, variable X-rays will still be observed as in narrow-line Seyfert 1 galaxies (panel a, Figure \ref{fig:cartoon}).

%

\subsection{Outflows at different scales}
Outflow is ubiquitous in simulations and observations of super-Eddington systems. But different types of objects reveal outflows on different scales.
On the scale of the corona, ultra-fast outflows \citep[e.g.,][]{Chartas:2002,Pounds:2003,Tombesi:2010,Nardini+2015} are winds with velocities that are an appreciable fraction of the speed of light. These must emerge from very close to the central black hole. On slightly larger scales, broad absorption lines \citep[e.g.,][]{Weyman+1991} and warm absorbers \citep[e.g.,][]{Halpern+1984} are each interpreted as energetic outflows, and are seen routinely (but not uniquely) in sources thought to have high Eddington ratio. Likewise, the inner broad line region, and particularly \ion{C}{iv}, often shows significant blue-shifts that have been linked to outflow on broad-line region scales \citep{Gaskell:1982,Richards:2002,BaskinLaor:2005,Trefoloni+2023}. \ion{C}{iv} weakness and blueshift are part of eigenvector 1, and also linked with X-ray weakness \citep[e.g.,][]{Richards+2001,Lusso+2021}.

Evidence for relatively large-scale outflow (pc-scale or beyond) is found from both ionized gas \citep[e.g.,][]{Harrison:2014} and molecular gas \citep[see review in][]{Veilleux:2020}. Ionized gas emission lines, such as the strong and ubiquitous [\ion{O}{iii}]$\lambda 5008$, often carry a prominent blue wing that is seen in narrow-line Seyfert 1 galaxies, ERQ/HotDOGS, FeLoBALs, and early very luminous quasars alike \citep[e.g.,][]{Heckman:1981,ZakamskaGreene2014,ShenHo2014,Liu+2026}. It is very hard to understand either the actual spatial scale or the energetics of these winds. It is also hard to know exactly how they are driven or how they may couple to winds or outflows on smaller scales. 
In some special cases, both nuclear UFOs and kpc-scale winds have been detected in the same source, allowing an exploration of how the energy is carried outward by the wind. Evidence for both energy-conserving and momentum-conserving winds has been seen \citep[see][and references therein]{Xu+2026}. JWST recently discovered blueshifted hydrocarbon dust absorption in the Hot DOG W2246-0526, which reveals dust outflowing at $\sim 5250$ km/s on $\sim 15-500$ parsec scales with a rate of $\sim 10^2-10^4\msun$ per year. One candidate outflow driving mechanism is the infrared radiation pressure on the dust grains, which can be locally super-Eddington due to enhanced dust opacity \citep{Vayner+2025}.

It is not straightforward to compare the energetic properties of the outflows produced by simulations with what is inferred from observations. The outflow velocity is best constrained observationally. For the hot phase, the projected speed follows from the blueshift of resonant absorption lines, \ion{Fe}{xxv} and \ion{Fe}{xxvi} for the UFOs in AGNs and the \ion{O}{viii}, \ion{Ne}{x} and Fe-L complexes for the cooler ULX winds \citep{Pinto+2016,Parker+2017}. 
The estimated velocity for the hot outflow produced from the inner region is around $\sim 0.1-0.3c$, which is quite consistent with the outflow velocity produced by the simulations. Fitting the profiles of these lines can return the column density and the ionization parameter observationally, but not the launch radius $r$. Since $r$ is needed to derive the mass flux and kinetic energy carried by the wind, those estimates are very uncertain and usually can only be estimated within a range. 

FeLoBALs are the one class that can break this degeneracy, because the ratios of excited-state to resonance-state \ion{Fe}{ii} and \ion{Si}{ii} column densities measure electron number density directly. Combined with the derived ionization parameter and the ionization luminosity from the AGN, the distance can be reasonably estimated. The resulting outflows are found between $\sim 1$ pc and $\sim 25$ kpc with kinetic energy flux reaching a few percent of the bolometric luminosity \citep{Choi+2022}. 
These estimates refer to very different scales, and this must be kept in mind when they are compared with the simulations. FeLoBAL radii correspond to $10^5-10^9\ r_g$ for a $10^8\msun$ black hole, far outside the range covered by the calculations in Figure \ref{fig:kin_efficiency}. 

The mass flux carried by the UFOs can typically reach 
$\sim 0.01-1\msun$ yr$^{-1}$, which is $\approx 5-10\%$ of the mass accretion rate. For outflows in broad absorption line quasars, the mass flux is also estimated to be a few percent of the mass accretion rate. The kinetic energy flux carried by the outflow is typically estimated to be $\sim 0.1\%-5\%$ of the bolometric luminosity \citep{Tombesi+2012,Fiore+2017,Sharma+2025}. These properties are quite consistent with the outflows produced by the simulations when the accretion rate is near the Eddington value (Figure \ref{fig:kin_efficiency}). For molecular outflow and other ionized winds, the inferred mass flux carried by the outflow can be $2-3$ orders of magnitude larger while the kinetic energy flux is still comparable due to lower outflow velocities. These outflows are outside the spatial scale studied by the simulations. 

\subsubsection{Clumpy outflow}
\label{sec:clumpy_wind}
X-ray observations of the winds in ULX systems have found large amplitude variability on timescales of tens to thousands of seconds \citep{Middleton+2011}. 
A multi-year XMM-Newton campaign \citep{Pinto+2026} confirms that the outflow in NGC 55 ULX-1 shows multi-phase winds with multiple velocity components. The broad absorption trough of the luminous quasar PDS 456 turns out to have five discrete velocity components reaching $20\%-30\%$ of the speed of light when the spectrum is resolved by XRISM \citep{Xrism2025}. This is very close to the outflow velocity typically found by numerical simulations of super-Eddington accretion disks. Furthermore, these observations suggest that the outflow is likely clumpy and highly inhomogeneous. It can consist of up to a million clumps of size $(0.2-1.2)\times 10^{15}$ cm located $(2-4) \times 10^{16}$ cm from the X-ray source, which is a few hundreds $r_g$ away from the black hole. Each individual clump is thought to have optical depth $\tau\sim 1$ so that these clumps can cause time variable obscuration of the X-rays. 

Early two-dimensional radiation MHD simulations suggest that the wind from super-Eddington accretion disks is indeed inhomogeneous \citep{Takeuchi+2013}. 
These clumps only show up around $\sim 500 r_g$ and the resolution is high enough to resolve the scale of $\sim 20\ r_g$ for one optical depth across the clump. It is suggested that the radiation Rayleigh-Taylor--like instability is responsible for the formation of these clumps \citep{Takeuchi+2014}. Three dimensional radiation hydro simulations just focusing on the photosphere of the wind region also find similar clumps \citep{Kobayashi+2018}. However, detailed quantitative comparison between the simulations and observations is still missing. Direct 3D radiation MHD simulations that cover the central engine as well as the region around $500\ r_g$ with enough resolution to resolve these small structures will be needed to provide a complete physical picture of these outflows. 

\subsection{LRD Modeling}
\label{sec:lrdmodel}

There are many papers that attempt to reproduce the spectra of the Little Red Dots, particularly in the case of accretion onto a central black hole. Typically, these papers are focused on radiative transfer, and so they posit gas densities and kinematics that may allow for the observed spectral features \citep[e.g.,][]{InayoshiMaiolino:2025,Chang+2026, Sneppen+2026}. A smaller number of papers aim to self-consistently model the gas distribution in the context of some assumed accretion model \citep{Liu+2025superedd,Begelman+2026}. Motivated by the empirical observed properties, there is some convergence that, compared with classical AGN, the LRDs have a denser and more massive gas cocoon or envelope, with some quasi-spherical geometry and high covering factor, and likely with a slow net outflow that is motivated by the high fraction of sources with observed P-Cygni profiles in their Balmer absorption \citep[e.g.,][]{Naidu+2026,Sneppen+2026}.

Taking a broad view, many of the papers posit that much of the emission from LRDs likely emerges from what we call the meso-scale (0.1 pc, or $\sim 10^6\ r_g$). The inner regions of the disk, within $r < 1000 \, r_g$, are likely obscured from our view by this cocoon/wind/envelope. Therefore, an exciting prospect for progress in the near-term is that more simulation work at these larger scales will yield growing insight into the observations of LRDs. Indeed, from our initial exploration, it seems possible that the accretion rate could be very super-Eddington on the meso-scale. Unlike the case for most accreting black holes, we may not be seeing a steady-state disk structure at all.

\section{Future directions}
Future improvements of our understanding of super-Eddington sources require joint effort between simulations and observations. Most existing simulations focus on the inner regions and start with idealized initial conditions while observations in the UV/Optical bands provide information at much larger scales. Extending simulation domains to cover a larger dynamical range is also possible now given improved computational power and more efficient numerical algorithms. Meanwhile, more physical processes need to be included in the simulations in order to produce the observables that can be directly compared with observations. Examples include self-gravity and dust physics, as well as line transport so that large scale outflows can be properly produced. 

Next generation instruments, such as GRAVITY+, are also pushing the limits of spatial resolution so that the mesoscale can be resolved. These new observations can put strong constraints on the kinematics of the flow at the sub-pc scale. Tying these scales with existing energetics measurements on kpc scales may help us work out how these large-scale winds are driven. At the same time, between GRAVITY+ and extremely large telescopes, our prospects are excellent to dramatically improve our handle on the black hole masses of quasars at cosmic noon, allowing for more detailed identification of super-Eddington sources for follow-up.

With the advent of the Legacy Survey of Space and Time (LSST), there will be considerable long-term light curves available to study the chromatic behavior of the optical/UV part of the disk with time \citep[e.g.,][]{Secunda:2023}. Using the chromatic behavior of disks discussed in \S \ref{sec:sim_variability}, we will have much better statistics on how the outer parts of disks behave as a function of optical luminosity and accretion rate. Furthermore, we may identify extreme sources turning on and off, as in \emph{1ES 1927+654}, to watch the phenomenology change in real time.

As outflows are unavoidable outcomes for super-Eddington accretion, calculating the properties of outflow from the mesoscales will be one important task for future simulations. This will be crucial to determine the broad band spectra for all super-Eddington sources and understand how the lines that are commonly used in observations are formed in the system. 
This is also one of the biggest uncertainties in all the models for LRDs, as the adopted density profiles as well as the kinematic structures may not be accurate. 
Then it becomes critical to compare these simulations with observations, including the continuum shape and line profiles, to test the whole picture of super-Eddington accretion for LRDs. 


Another major outstanding question is still the origin of supermassive black holes, and super-Eddington accretion may well play a central role. At this moment, we have tantalizing hints that LRDs may be a sign of very high accretion rates at early times, but no concrete handle on their black hole masses. More generally, the verdict is still out on whether early black holes are generally overmassive \citep[e.g.,][]{Maiolino+2024}. If so, it is tempting to invoke early super-Eddington growth to grow so much black hole mass so quickly \citep[e.g.,][]{Trinca+2024}. In the coming decade, we will make progress on the role of super-Eddington growth in black hole seeding by pushing the redshift frontier beyond $z \sim 8$ with \emph{JWST} and the E-ELT, as well as hopefully by determining robust black hole masses for sources in the first billion years. 

\section*{Acknowledgements}
During the preparation of this review, the authors used Claude to collect the data that were used for Figures~\ref{fig:redshift_lum}, \ref{fig:efficiency} and \ref{fig:kin_efficiency}. After using these tools, the authors reviewed and verified the data and take full responsibility for the accuracy of the publication.  

We are grateful to many inspiring discussions and comments from our colleagues when we prepare the review, including (in alphabetical order): Xiaohui Fan, Feng Hua, Karen Leighly, Yilun Ma, Roberto Soria, James Stone. 

Part of this review is based upon work supported by the National Science Foundation under Grant No. 2306950.

\bibliographystyle{ar-style2}
\bibliography{citations}

\clearpage
\section{Supplement}

\subsection{Bolometric Corrections}
\label{sec:BC}

In order to compare between samples, and with theory, it is useful to estimate the full bolometric output of our sample, as summarized in Figure \ref{fig:redshift_lum}. In Figure \ref{fig:sed_cartoon}, we show that the broad-band SEDs of most flavors of super-Eddington AGN are quite distinct from standard AGN. These differences make a determination of the bolometric luminosity particularly challenging. Here we discuss the bolometric corrections that we adopted in creating Figure \ref{fig:redshift_lum}, with the caveat that significant uncertainty remains.

For narrow-line Seyfert 1 galaxies, the bolometric luminosity is derived using the rest frame optical monochromatic luminosity at $5100\AA$ as $L_{\rm bol}=9.26L_{5100}$. 
The value $9.26$ is the mean SED correction of \cite{Richards+2006}, which is also adopted by \cite{Shen+2011}. Notice that this is based on the mean SED of all type 1 AGNs independent of luminosity, mass and accretion rate. Based on accretion flow models for narrow-line Seyfert 1 galaxies, \emph{optxagnf} \citep{Done+2012} and its successor \emph{qsosed} \citep{KubotaDone2018} predict a bolometric correction that varies with black hole mass and Eddington ratio, because the SED peak shifts to shorter wavelengths as the Eddington ratio increases, which will cause the bolometric correction factor anchored at $5100\AA$ to be larger. \cite{Azadi+2025} derive the correction factor varying from $2.5$ to $6.6$ with the median value of $3.4$, which is a factor of $2-3$ smaller than what we adopt. 

WLQs are commonly observed in the rest frame UV and the bolometric luminosity is scaled with the monochromatic luminosity at $3000\AA$ as 
$L_{\rm bol}=5.15L_{3000}$. The correction factor $5.15$ is also based on the mean SED of type 1 AGNs \citep{Richards+2006,Shen+2011}. For ERQs, we adopt the bolometric correction factor given by \cite{Runnoe+2012} as $L_{\rm bol}=8L_{3\mu m}$ anchored by the monochromatic luminosity at $3\mu m$. This is also consistent with the infrared correction factor given by \cite{Richards+2006}. However, this factor is just a crude estimate as the underlying disk structures for ERQs can be very different from typical type 1 AGNs. For the hyperluminous infrared selected Hot DOGs, the correction factor is chosen to be unity.

For all three LRD samples we adopt the LRD-specific bolometric correction of \cite{Greene+2025}, $L_{\rm bol}=5.4\,L_{5100}$, where $L_{5100}$ is the \emph{observed}, not dust-corrected, monochromatic luminosity at rest-frame $5100\AA$. This correction is measured from the X-ray to far-infrared SEDs of two luminous LRDs, in which more than half of the bolometric luminosity emerges in the rest-frame optical, and it is roughly half the value for type-1 AGNs. For the spectroscopic LRDs, the 146-object NIRSpec/PRISM sample of \cite{deGraaff+2026} and the 27 low-redshift DESI LRDs of \cite{Lin+2026}, $L_{5100}$ is measured directly from the spectra. For the photometric sample of \cite{Akins+2025}, we follow \cite{Greene+2025} and derive $L_{5100}$ from a power-law interpolation of the NIRCam/MIRI photometry bracketing rest-frame $5100\AA$. The resulting bolometric luminosities are a median of $1.6$ dex below the values given by \cite{Akins+2025}, who applied the type-1 correction $L_{\rm bol}=5.15L_{3000}$ to an intrinsic, dust-corrected SED fit. With the common correction, the photometric and spectroscopic LRDs at $z>2$ occupy the same region in Figure \ref{fig:redshift_lum}, with median $\log L_{\rm bol}=44.3$ and $44.2$, respectively.

\subsection{Summary Table of Sample Properties}

\begin{table}[ht]
\tabcolsep4pt
\caption{Samples shown in Figure~\ref{fig:redshift_lum}. For each class we
list the rest-frame band to which the luminosity is anchored, the
median of that monochromatic luminosity
$L_{\rm obs}=\nu L_\nu$ (in erg\,s$^{-1}$), the bolometric correction used
to place the objects on the $L_{\rm bol}$--$z$ plane, and the median
redshift. The three LRD samples share the bolometric correction of
\citet{Greene+2025} but are listed separately because their $L_{5100}$ is
measured differently. This table and the underlying per-object
lists are available as machine-readable tables in the Supplemental Material.}
\label{tab:zLsamples}
\begin{center}
\footnotesize
\begin{tabular}{@{}lccccc@{}}
\hline
Sample & $N$ & Observed band & Median & Bolometric & Median $z$ \\
 & & (rest frame) & $\log_{10}L_{\rm obs}$ & correction ($\to L_{\rm bol}$) & \\
\hline
NLS1 &  42 & 5100\,\AA            & 44.09 & $\times 9.26$                      & 0.09 \\
Weak-line quasars & 137 & 3000\,\AA        & 46.19 & $\times 5.15$                      & 2.03 \\
ERQ         &    97 & 3\,$\mu$m            & 45.71 & $\times 8$                         & 2.50 \\
Hot DOG$^{\rm c}$ &    20 & IR 8--1000\,$\mu$m   & 47.50 & $L_{\rm bol}/L_{\rm IR}$, median $1.46$ & 3.19 \\
LRD, photometric$^{\rm a}$ & 434 & 5100\,\AA            & 43.60 & $\times 5.4$                       & 6.51 \\
LRD, spectroscopic$^{\rm b}$ & 146 & 5100\,\AA            & 43.46 & $\times 5.4$                       & 4.98 \\
LRD, low-$z$ (DESI)$^{\rm b}$ &  27 & 5100\,\AA            & 43.59 & $\times 5.4$                       & 0.47 \\
\hline
\end{tabular}
\end{center}
\begin{tabnote}
$^{\rm a}$For the photometric LRDs of \citet{Akins+2025}, $L_{5100}$ is the
observed (not dust-corrected) luminosity interpolated from the NIRCam/MIRI
photometry bracketing rest-frame 5100\,\AA, and the correction is that of
\citet{Greene+2025}.
$^{\rm b}$For the spectroscopic LRDs of \citet{deGraaff+2026} and \citet{Lin+2026}, $L_{5100}$ is measured from the spectra, and the correction is that of \citet{Greene+2025}.
$^{\rm c}$The Hot DOGs are the one class for which no single correction is
assumed: both $L_{\rm bol}$ and $L_{\rm IR}$ are taken object by object from
the SED fits of \citet{Tsai+2015}, and the implied ratio runs from 1.25 to 2.37
(median 1.46).
\end{tabnote}
\end{table}

\end{document}